\documentclass[12pt,preprint]{aastex}

\shorttitle{Hot and Diffuse Clouds near GC} \shortauthors{Oka et
al.}

\begin{document}


\title{Hot and Diffuse Clouds near the Galactic Center \\
Probed by Metastable H$_3^+$\altaffilmark{1, 2, 3}}


\altaffiltext{1}{Based in part on observations obtained at the
Gemini Observatory, which is operated by the Association of
Universities for Research in Astronomy, Inc., under a cooperative
agreement with the NSF on behalf of the Gemini partnership: the
National Science Foundation (United States), the Particle Physics
and Astronomy Research Council (United Kingdom), the National
Research Council (Canada), CONICYT (Chile), the Australian Research
Council (Australia), CNPq (Brazil) and CONICET (Argentina).}

\altaffiltext{2}{Based on data collected at the United Kingdom
Infrared Telescope, which is operated by the Joint Astronomy Centre
on behalf of the U. K. Particle Physics and Astronomy Research
Council.}

\altaffiltext{3}{Based on data collected at the Subaru Telescope,
which is operated by the National Astronomical Observatory of
Japan.}
\author{Takeshi Oka\altaffilmark{4}, T. R. Geballe\altaffilmark{5}, Miwa
Goto\altaffilmark{6},\\
Tomonori Usuda\altaffilmark{7}, and Benjamin J.
 McCall\altaffilmark{8}}

\email{t-oka@uchicago.edu}


\altaffiltext{4}{ Department of Astronomy and Astrophysics, and
Department of Chemistry, The Enrico Fermi Institute, University of
Chicago, Chicago, IL  60637 USA } \altaffiltext{5}{Gemini
Observatory, Hilo, Hawaii  96720  USA} \altaffiltext{6}{Max Planck
Institute for Astronomy,  Heidelberg, Germany}
\altaffiltext{7}{Subaru Telescope, National Astronomical Observatory
of Japan, Hilo, Hawaii 96720 USA} \altaffiltext{8}{ Department of
Chemistry and Department of Astronomy, University of Illinois
Urbana-Champaign, Urbana, IL  61801-3792  USA}


\begin{abstract}

Using an absorption line from the metastable ($J$, $K$) = (3, 3)
level of H$_3^+$ together with other lines of H$_3^+$ and CO
observed along several sightlines, we have discovered a vast amount
of high temperature ($T$ $\sim$ 250 K) and low density ($n$ $\sim$
100 cm$^{-3}$) gas with a large velocity dispersion in the Central
Molecular Zone (CMZ) of the Galaxy, i.e., within 200 pc of the
center. Approximately three-fourths of the H$_3^+$ along the line of
sight to the brightest source we observed, the Quintuplet object GCS
3-2, is inferred to be in the CMZ, with the remaining H$_3^+$
located in intervening spiral arms.  About half of H$_3^+$ in the
CMZ has velocities near $\sim$ $-$100 km s$^{-1}$ indicating that it
is associated with the 180 pc radius Expanding Molecular Ring which
approximately forms outer boundary of the CMZ. The other half, with
velocities of $\sim$ $-$50 km s$^{-1}$ and $\sim$ 0 km s$^{-1}$, is
probably closer to the center. CO is not very abundant in these
clouds. Hot and diffuse gas in which the (3, 3) level is populated
was not detected toward several dense clouds and diffuse clouds in
the Galactic disk where large column densities of colder H$_3^+$
have been reported previously. Thus the newly discovered environment
appears to be unique to the CMZ. The large observed H$_3^+$ column
densities in the CMZ suggests an ionization rate much higher than in
the diffuse interstellar medium in the Galactic disk. Our finding
that the H$_3^+$ in the CMZ is almost entirely in diffuse clouds
indicates that the reported volume filling factor ($f$ $\geq$ 0.1)
for $n \geq$ 10$^{4}$ cm$^{-3}$ clouds in the CMZ is an overestimate
by at least an order of magnitude.

\end{abstract}



\section{Introduction}

The Central Molecular Zone (CMZ, also called the nuclear molecular
disk) of the Galaxy, a region of radius $\sim$ 200pc, is currently
considered to contain high density ($n$ $\geq 10^4$ cm$^{-3}$)
molecular gas with a high volume filling factor \citep{mor96}.  A
significant fraction of the molecular gas has unusually high
temperature ($\sim$ 200 K) and large velocity dispersion
(15--50~km~s$^{-1}$), indicative of the highly energetic, turbulent
nature of the CMZ. Unlike the hot spots in the Galactic disk where a
luminous star heats nearby dust, which in turn heats the associated
gas, the hot gas in the CMZ extends over large distances and
suggests a direct and widespread gas heating mechanism. On the other
hand, dust in the same area has been observed to have low
temperatures ($\leq$ 30 K) from far-infrared \citep{ode84} and
submillimeter \citep{pie00} measurements. Studies of the hot gas in
the CMZ provide information vital for understanding the unusual
activity near the nucleus of the Galaxy, which includes non-thermal
magnetic phenomena \citep{yus84}, extended X-ray emission
\citep{koy89, koy96} and their interactions with molecular clouds
\citep{lis85, tsu97, oka01, yus02}.

The most direct evidence for high temperature gas in the CMZ comes
from radio observations of molecules in high rotational levels.
\citet{wil82} observed the inversion spectrum of NH$_3$ in
absorption up to the ($J$, $K$) = (8, 8) and (9, 9) metastable
levels, 687 K and 853 K above the ground level, respectively, toward
Sgr B2 and estimated a temperature of $T$ = 200 K and cloud density
of $n$ = 10$^4$ cm$^{-3}$. \citet{mau86} and \citet{hut93} observed
other clouds in the CMZ, finding NH$_3$ emission lines from the (7,
7) and (6, 6) metastable levels, respectively, and reporting similar
temperatures and densities. Recently, \citet{her02} obtained similar
results toward Sgr A. \citet{har85}, using the $J = 7-6$
submillimeter emission line of CO found 10$^4$ M$_{\odot}$ of warm
dense gas in the central 10 pc of the Galaxy with $T$ $\sim$ 300 K
and $n$ = 3 $\times$ $10^4$ cm$^{-3}$.  \citet{kim02}, using the
same transition, found a lower mean density and temperature over a
much larger volume centered on Sgr~A. The $J = 7-6$ transition has a
critical density of $n_{crit} \geq 10^6$ cm$^{-3}$ \citep{gre78} and
thus probes dense regions of the CMZ. Recently, \citet{rod01} used
\emph{ISO} measurements of the $S$(0) - $S$(5) rotational emission
lines of H$_2$ to investigate 16 molecular clouds within the central
$\sim$ 500 pc of the Galaxy. Since critical densities of those
transitions are lower, these lines can be detected in low density
clouds if a sufficient column density of H$_2$ exists.  They
reported $T$ = 150 $-$ 600 K and $n$ = 10$^{3.5-4.0}$ cm$^{-3}$,
similar to other studies. Very recently, \citet{rod04} used
observations of atomic fine structure lines, believed to arise in
photodissociation regions associated with these clouds, to derive
densities of $\sim$ 10$^3$ cm$^{-3}$.

Here we report the discovery of extensive \emph{low} density ($n$
$\sim$ 100 cm$^{-3}$) and high temperature ($T$ $\sim$ 250 K)
molecular gas in the CMZ using infrared absorption lines of H$_3^+$,
including a line from the ($J$, $K$) = (3, 3) metastable level,
which was first observed toward the Galactic center by
\citet{got02}. H$_3^+$ (protonated H$_2$) is a unique astrophysical
probe in that it is observed with comparable column densities ($\sim
10^{14}$ cm$^{-2}$) in dense cloud cores \citep{geb96, mcc99, bri04}
and diffuse clouds \citep{mcc98a, geb99, mcc02}. Contrary to the
expectation from chemical model calculations which predict orders of
magnitude lower H$_3^+$ number density for diffuse clouds (because
of rapid dissociative recombination with electrons), our
observations have established that column densities of H$_3^+$ per
unit visual extinction typically are an order of magnitude
\emph{higher} in diffuse clouds ($N$(H$_3^+$) $\sim 4.4 \times
10^{13}$ cm$^{-2}$ $A_V$) than in dense cloud cores ($N$(H$_3^+$)
$\sim 3.6 \times 10^{12}$ cm$^{-2}$ $A_V$) \citep{mcc99, mcc02,
oka04b}. Since the $A_V$ of 25 - 40 mag \citep{cot00} toward the
Galactic center arises mainly in diffuse clouds \citep{wil79, but86,
pen94, whi97}, H$_3^+$ is a readily available tool for studying
sightlines toward the Galactic center. Indeed the observed column
densities of $N$(H$_3^+$) $\sim (3 - 5) \times 10^{15}$ cm$^{-2}$
toward \objectname{GCS 3-2} and \objectname{GC IRS 3} \citep{geb99,
got02} are an order of magnitude higher than other sightlines in the
Galactic disk. H$_3^+$ is also important for the study of the CMZ
since, being a charged molecule, it is sensitive to ionization of
molecular gas which is expected to be efficient in the CMZ with its
intense X-ray sources and high magnetohydrodynamic activity
\citep{mor96}.

H$_3^+$ has the structure of an equilateral triangle like the
hydrogen atoms in NH$_3$, but being simpler, lighter, and
electrically charged, it has three salient features not found in
NH$_3$. First, H$_3^+$ is planar. Thus totally symmetric rotational
levels, e.g., (0, 0), (2, 0), etc., are not allowed by the Pauli
exclusion principle. This makes the (1, 1) level the rotational
ground state. Second, the metastability of H$_3^+$, although similar
to that of NH$_3$ qualitatively, is very different quantitatively.
The crucial difference relevant to this paper is that while NH$_3$
in the (2, 2) level takes $\sim$ 200 years to decay to the (1, 1)
level by spontaneous emission \citep{oka71}, the same emission in
H$_3^+$ takes only 27 days.  On the other hand, spontaneous emission
from the (3, 3) level of H$_3^+$ is rigorously forbidden by the
dipole selection rules and the absence of the (2, 0) level
\citep{pan86}. This results in a highly non-thermal population
distribution in the H$_3^+$ rotational levels in low density gas.
More details can be found in \citet{oke04} (hereafter Paper I). Here
we simply emphasize that the spontaneous emission time of the (2, 2)
$\rightarrow$ (1, 1) transition, 2.35 $\times 10^6$ s, is a very
reliable number with an uncertainty of at most 5 \% based on the ab
initio calculation by \citet{nea96}, and it sets an accurate time
standard for the thermalization of interstellar H$_3^+$. Spontaneous
emissions from higher levels such as (2, 1) $\rightarrow$ (2, 2) (20
days), (3, 2) $\rightarrow$ (2, 1) (16 hours), and (3, 1)
$\rightarrow$ (2, 2) (8 hours) have even shorter lifetimes. The
spontaneous emission rate increases rapidly with increasing $J$ and
$K$ and thus rotationally hot H$_3^+$ rapidly cools except when in
metastable levels, (4, 4), (5, 5), (6, 6) etc. This has implications
even for laboratory experiments \citep{kre02, kre04}.

Third, collisions between H$_3^+$ and H$_2$ are qualitatively
different from those between NH$_3$ and H$_2$. They actually are
chemical reactions H$_3^+$ + H$_2$ $\rightarrow$ (H$_5^+$)$^*$
$\rightarrow$ H$_3^+$ + H$_2$ in which the protons may scramble in
the activated complex (H$_5^+$)$^*$ during its short lifetime
\citep{Oka04a}. Therefore, unlike NH$_3$ or other neutral molecules
such as H$_2$O and H$_2$CO, ortho- and para- nuclear spin
modifications are efficiently converted to each other by collisions
with H$_2$ \citep{cor00}. In addition, a collision between H$_3^+$
and H$_2$ has higher cross section than one between NH$_3$ and
H$_2$, because of the long range $r^{-4}$ Langevin potential, making
critical densities lower. Assuming that the collision rate constant
between H$_3^+$ and H$_2$ is close to the Langevin rate constant,
$k_L$ = 2$\pi$e($\alpha$/$\mu$)$^{1/2}$ = 1.9 $\times 10^{-9}$
cm$^3$ s$^{-1}$, where $\alpha$ is the polarizability of H$_2$ and
$\mu$ is the reduced mass of H$_3^+$ and H$_2$ \citep{rid04}, we
find that the critical density for the (2, 2)$\rightarrow$ (1, 1)
spontaneous emission is $\sim$ 200 cm$^{-3}$, comparable to the
densities of diffuse clouds. Therefore, an absence of H$_3^+$ in the
(2, 2) state in warm gas is definitive evidence for low density. At
low temperatures the rate for the collision-induced transition from
(1, 1) to (2, 2) is much less than the Langevin rate because of the
principle of detailed balancing. The value is also lower at very
high temperatures because of dilution by collisional transitions
from (1, 1) to other levels higher than (2, 2), but this is
compensated by the channeling via rapid spontaneous emission of most
of para-H$_3^+$ in high rotational levels to the (2, 2)
$\rightarrow$ (1, 1) transition. These problems are taken into
account in the model calculations of Paper I.

\section{Observations}

Three spectrometers on three telescopes were used to make the
measurements presented here; the observing log is given in Table 1.
The Phoenix Spectrometer mounted on the 8~m Gemini South Telescope
is the most suitable for this study because of its location in the
southern hemisphere and its high spectral resolution (5 km
s$^{-1}$), but its wavelength coverage is limited by its narrow
spectral coverage and the availability of order-blocking filters.
The Infrared Camera and Spectrograph (IRCS) on the 8~m Subaru
Telescope is a powerful survey instrument because of its wide
wavelength coverage, although its resolution is relatively low (15
km s$^{-1}$). The Cold Grating Spectrometer 4 (CGS4) on the 3.8~m
United Kingdom Infrared Telescope (UKIRT) has a resolution of 8 km
s$^{-1}$ and can be tuned to any wavelength in the 1 $-$ 5 $\mu$m
interval, but like Phoenix has a narrow spectral coverage at each
tuning.  Details of operations of the spectrometers and data
reduction are given in \citet{mcc02}, \citet{got02} and
\citet{geb99}, respectively.

Targets were chosen from bright young infrared stars with magnitudes
$L$ $\leq 7.5$ which have clean infrared continua. Most of the
bright stars near the Galactic center are late-type stars whose
continua are not smooth due to photospheric atomic and molecular
absorptions and thus are problematic for detections of weak
interstellar absorption lines. Currently, out of the 50 bright
Galactic center objects studied by \citet{nag93}, only 6 have been
used. This difficulty could be partially overcome in the near future
by subtracting standard photospheric absorptions from observed
spectra. We have observed sightlines toward 8 infrared sources
listed in Table 1. The best set of spectrum has been obtained in the
sightline toward the brightest source GCS 3-2 and this spectrum is
discussed in detail below. The rest of 7 sightlines have been
observed with lower resolution and will be discussed briefly in
Section 3.2. More detailed studies on these sightlines will be
published in future papers.

Compared to radio observations, infrared absorption spectroscopy has
the disadvantage that mapping is infeasible because observable
sightlines are limited to directions of bright infrared sources.
Radio observations are also more sensitive and often with higher
spectral resolution. However, infrared spectroscopy has an advantage
in its inherent high spatial resolution. Moreover, the determination
of column densities via infrared spectroscopy is often more
straightforward because of lower optical depths and availability of
several transitions in the same wavelength region. In particular the
presence of spectral lines starting from qualitatively different
rotational levels, such as the (1, 1) ground level, (2, 2) unstable
level, and (3, 3) metastable level, makes H$_3^+$ a unique
astronomical probe for studies in this paper.

\section{Results}

Spectra of GCS 3-2 at the $R$(1, 1)$^l$ , $R$(3, 3)$^l$, and $R$(2,
2)$^l$ transitions of H$_3^+$ in the $L'$ window and the $R$(1)
transition of the $v$ = 2 - 0 overtone band of CO in the $K$ window
are shown in Fig.~1. See \citet{lin01} for the nomenclature of
$R$(1, 1)$^l$ etc. GCS 3-2 is a young massive star bright in the
infrared ($L$ = 3.03, $K$ = 6.19 \citep{nag93}). It is one of the
Quintuplet stars \citep{nag90, oku90} in the region with recent
massive star formation located approximately where the non-thermal
Radio Arc crosses the Galactic plane at the Galactic coordinates $l$
= 10$\arcmin$, $b$ = $-$4$\arcmin$, close to the compact thermal
structures known as the Pistol and the Sickle \citep{gen94, cot00}.

The intense and sharp absorptions at $-$52 km s$^{-1}$, $-$33 km
s$^{-1}$, $-$5 and +2 km s$^{-1}$, and +19 km s$^{-1}$ and +23 km
s$^{-1}$, clearly visible both in the H$_3^+$ $R$(1, 1)$^l$ and CO
$R(1)$ spectra of Fig.~1 have been known since the early days of the
21 cm H~I spectroscopy and radio-observations of OH, H$_2$CO, and CO
to be due to clouds in intervening spiral arms \citep{oor77}. The
$-$52 km s$^{-1}$ absorption arises in the 3 kpc arm \citep{van57,
rou60} and the $-$33 km s$^{-1}$ absorption is due to the 4.5 kpc
arm \citep{men70}. Absorptions near +20 km s$^{-1}$ have generally
been ascribed to the ``20 km s$^{-1}$ cloud" believed to lie within
the CMZ \citep{gus81a, geb89}

Absorption components near 0 km s$^{-1}$, which are ubiquitous and
dominant in H~I spectra, often are ascribed to ``local" arms within
a few kpc of the solar system. For CO and H$_3^+$, however, the
absorptions at these velocities may not be local because, unlike the
H~I line, they are not uniformly observed in all sightlines but only
toward the Galactic center. \citet{whi79} suggested that some of
them are near the Galactic center. Analyses and discussions of those
sharp lines corresponding to low temperature CO and H$_3^+$ in
intervening spiral arms are outside the scope of this paper and will
be given separately.

\subsection{Non-thermal Rotational Distribution Toward GCS 3-2}

Initially we focus our attention on the three broad absorption features
with large equivalent widths at $-$123 km s$^{-1}$, $-$97 km
s$^{-1}$ and $-$52 km s$^{-1}$ clearly visible in the $R$(3, 3)$^l$
spectrum in Fig.~1. These absorptions were present in our earlier
paper \citep{got02}, but with less clarity. In the new spectrum weak
and broad absorptions at lower velocities are also evident. We argue
in the following that the three prominent absorption features in the
(3, 3) line arise in high temperature, low density turbulent clouds
in the CMZ.

First, note that the $R$(3, 3)$^l$ spectrum does not show any of the
sharp features that are observed in the $R$(1, 1)$^l$ spectrum and
in the $R$(1) CO spectrum. Note also that the CO spectrum does not
show the broad features of the $R$(3, 3)$^l$ spectrum beyond
2$\sigma$ of 1~\%. This difference cannot be due to the slightly
lower resolution of the $R$(3, 3)$^l$ spectrum. Clearly for the most
part the metastable H$_3^+$ and CO do not coexist. CO is found in
relatively high density and low temperature areas with low velocity
dispersion mostly in the intervening spiral arms, while H$_3^+$ in
the metastable level exists in hotter, more turbulent regions in the
CMZ.

Second, note that the H$_3^+$ $R$(1, 1)$^l$ spectrum also shows the
broad absorptions at $-$123 km s$^{-1}$, $-$97 km s$^{-1}$ and $-$52
km s$^{-1}$, although the $-$52 km s$^{-1}$ component is overlapped
by the strong and sharp line contribution from cold H$_3^+$ in the 3
kpc arm. The strengths of the broad $R$(1, 1)$^l$ and $R$(3, 3)$^l$
absorptions at these velocities are approximately the same. Since
the strengths of the $R$(1, 1)$^l$ and $R$(3, 3)$^l$ transitions are
$\mid \mu_{ij}\mid^2$ = 0.01407 D$^2$ and 0.01914 D$^2$,
respectively, their equal intensities imply a population ratio $N$(3,
3)/$N$(1, 1) = (14/3) exp ($-$361/$T$$_{ex}$) = 0.735, corresponding
to $T$$_{ex}$ = 195 K. Because of spontaneous emission between
rotational levels of H$_3^+$, the kinetic temperature of the clouds
$T$ is at least as high as $T_{ex}$ and is likely much higher.

Third, note that the H$_3^+$ $R$(2, 2)$^l$ spectrum does not show
any absorption near $-$123 km s$^{-1}$ and $-$97 km s$^{-1}$
($\Delta$$I$ $\leq 0.3$ \%) and at best a very weak absorption near
$-$52 km s$^{-1}$. This demonstrates a remarkably non-thermal
H$_3^+$ rotational distribution in which the (3, 3) level 361 K
above the ground level is highly populated while the (2, 2) level
151 K above ground is scarcely populated. It also shows that that
the cloud density is considerably lower than the critical density
for the (2, 2) $\rightarrow$ (1, 1) spontaneous emission, $N_{crit}$
$\sim$ 200 cm$^{-3}$. Thus, the high temperature and low density of
the clouds are established.

The high negative velocities $\sim$$-$100 km s$^{-1}$ of the hot and
diffuse clouds, peaking at $-$123 km s$^{-1}$ and $-$97 km s$^{-1}$,
suggest that they are part of a large-scale structure moving away
from the Galactic center at high speed that was initially proposed
as the Expanding Molecular Ring (EMR) by \citet{kai72},
\citet{sco72} and \citet{kai74} from the ($l$-$V$) maps of OH,
H$_2$CO, and CO. It was later also interpreted as part of a
parallelogram by \citet{bin91} and as an Expanding Molecular Shell
(EMS) by \citet{sof95} based on the radio $^{13}$CO observations by
\citet{bal87, bal88}. \citet{sof95} estimated that the total CO
emission from the EMS amounts to 15 \% of the total emission from
the CMZ. From a large-scale CO survey of the Galactic center,
\citet{oka98a} pointed out the possibility that the EMR may consist
of multiple arm-features that could be associated with different
Lindblad resonances \citep{bin87} discussed by \citet{bin91} and
\citet{bli93}. The observed double peak of the $R$(3, 3)$^l$ and
$R$(1, 1)$^l$ spectra in this velocity interval must provide
information on this issue. It is interesting to note that some CO is
observed in the $-$97 km s$^{-1}$ cloud but not in the $-$123 km
s$^{-1}$ cloud. The small amount of CO in the cold and high density
areas of those clouds in the particular direction of GCS 3-2 ($l$ =
10' and $b$ = $-$4') (Fig. 1) is in agreement with the observation
of \citet{oka98a} (see their $l$-$V$ maps for $b$ = $-$4$\arcmin$ on
page 493).

The presence of $R$(3, 3)$^l$ absorption at $-$52 km s$^{-1}$ might
suggest that hot and diffuse gas also exists in the ``$-$50 km
s$^{-1}$ cloud" that has been ascribed to the 3 kpc arm. The narrow
$R$(1, 1)$^l$ and CO features at this velocity probably do arise in
the arm, which merges into the corotating stars in the Galactic
Stellar Bar at about 2.4 kpc \citep{bin91, wad94, mez96}. However,
the large velocity dispersion of the $R$(3, 3)$^l$ absorption
component and the warm gas temperature suggest that, like the higher
negative velocity $R$(3, 3)$^l$ absorptions, it originates in the
CMZ.

Comparison of the H$_3^+$ $R$(1, 1)$^l$ spectrum with the CO $R(1)$
spectrum shows that in the former there is a wide and shallow
absorption trough at velocities between about $-$40 km s$^{-1}$ and
$+$30 km s$^{-1}$, with a depth of about two percent of the
continuum. The existence of this feature was already noted in our
earlier work \citep{geb99, got02}. In contrast, the CO spectrum
returns to the continuum level between the narrow absorptions at
$-$52, $-$32, $-$5, and $+$19 km s$^{-1}$. Separation of the trough
and sharp lines in the $R$(1, 1)$^l$ spectrum is discussed in more
detail below in Section 4.1 (see Fig.~4 given later). The H$_3^+$
absorption trough is also present, but weaker in the $R$(3, 3)$^l$
spectrum and thus demonstrates the presence of diffuse clouds with
little CO at somewhat lower temperatures than the gas in the EMR. It
is difficult to determine the number and natures of the lower
velocity warm clouds that contribute to the trough, because of their
weak absorptions and the contamination by strong and narrow spectral
features. However, some of the absorption is certainly due to the
aforementioned 20 km s$^{-1}$ cloud and it seems probable that the
remainder also is located inside the CMZ; the high velocity spread
is in sharp contrast to that of H$_3^+$ in many cold dense and
diffuse clouds in the Galactic disk which always give narrow lines
\citep{geb96, mcc98a, geb99, mcc99, mcc02, mcc03, bri04}.
Observations of many more sources distributed over a wider region of
the CMZ under high spectroscopic resolution will be attempted in the
near future to obtain a more detailed picture of those clouds.

\subsection{Spectra of Other Infrared Sources in the CMZ}

The non-thermal distribution in the lower rotational levels of
H$_3^+$ has also been observed toward other infrared sources in the
CMZ. Fig.~2 shows Subaru spectra of the $R$(1, 1)$^l$, $R$(3, 3)$^l$
and $R$(2, 2)$^l$ lines toward eight infrared sources. \objectname{GC IRS 1W}, GC
IRS 3, and \objectname{GC IRS 21} are near Sgr A* \citep{bec78, tol89} at
Galactic coordinates $l$ = 359.94$^{\circ}$ and $b$ =
$-$0.05$^{\circ}$, while GCS 3-2, \objectname{NHS 21}, \objectname{NHS 22}, \objectname{NHS 25}, and \objectname{NHS 42}
are at nearly the same Galactic latitude but with higher longitudes
ranging from $l$ = 0.10$^{\circ}$ to 0.16$^{\circ}$ \citep{nag90}.
The locations of these sources are shown in Fig.~3 superimposed on
the VLA 20 cm image of the Galactic center by \citet{yus87}, which
shows the high MHD activity in the region with both thermal and
non-thermal emission.

All targets in Fig.~2 show absorption in the metastable $R$(3,
3)$^l$ line, the fingerprint of high temperature, and the absence or
at best very weak absorption in the $R$(2, 2)$^l$, signifying low
density when the $R$(3, 3)$^l$ line is present. The velocities of
the absorption features vary greatly from source to source. As the
sightlines are separated by 1$\arcmin$ -- 12$\arcmin$, this provides
supporting evidence that the absorbing clouds are within the CMZ and
not in spiral arms farther from the Galactic center. Thus, these
observations strongly suggest that hot and diffuse gas exists widely
in the CMZ.

Of particular interest in Fig.~2 is the pronounced broad $R$(3,
3)$^l$ absorption at v$_{LSR}$ $\sim$ $+$50 km s$^{-1}$ in the
spectrum of GC IRS 3, which also was noted by \citet{got02}.  The
hot and diffuse gas producing this absorption apparently is a
component of the ``50 km s$^{-1}$ cloud", a complex of giant
molecular clouds within 10 pc of the Galactic nuclei \citep{gus81}
that is a key to understanding Sgr A and its environment
\citep{bro84, lis85}.  From measurements of NH$_3$ emission and
H$_2$CO absorption, \citet{gus81} and \citet{gus83} concluded that
the 50 km s$^{-1}$ cloud is sandwiched between Sgr A West in the
front and Sgr East in the back.  This is clearly confirmed in the
recent observations of CS by \citet{lan02} and the measurements of
H~I by \citet{lan04}. Hence GC IRS 3 must be behind the 50 km
s$^{-1}$ cloud and at least somewhat behind the Galactic nucleus.
The 50 km s$^{-1}$ component is not observed toward GC IRS 1W,
indicating that this source is close to or in front of the nucleus.

\citet{geb89} saw similar behavior of the velocity components in
fundamental band CO lines; the 50 km s$^{-1}$ component was observed
toward GC IRS 3 and GC IRS 7, but not toward GC IRS 1 and GC IRS 2.
They suggested, in view of the close proximity of these four sources
on the plane of the sky, that the CO absorption at 50 km s$^{-1}$
might not be due to the 50 km s$^{-1}$ cloud but rather to the inner
edge of the circumnuclear molecular ring orbiting the nucleus at a
radius of $\sim$ 2 pc \citep{gen94}. The present observations,
however, indicate that the absorption is likely due to the 50 km
s$^{-1}$ cloud since one would not expect to detect such a large
column density of H$_3^+$ in the circumnuclear molecular ring.

Detailed analyses of the individual spectra in Fig. 2 together with many
more spectral lines observed at Subaru will be given in a separate paper.
Here we simply emphasize that the hot and diffuse clouds are distributed
widely in the CMZ. The large variations in velocity and density of clouds
within 0.2$^{\circ}$ of the center is not in disaccord with the CO
observations by \citet{oka98a} (see their $l$-$V$ maps on page 493). Large
spatial variations in dust absorption have also been reported by
\citet{ada04}.

\subsection{Spectra of H$_3^+$ in the Galactic disk}

We have attempted to detect the $R$(3, 3)$^l$ line toward several
dense and diffuse clouds where large column densities of H$_3^+$
have previously been observed in the low-lying (1, 1) and (1, 0)
levels, in order to determine if cold dense or diffuse clouds are
surrounded by diffuse and warmer gas. We have not been able to
detect the line in any of AFGL 2136 and W 33A \citep{geb96}, Cygnus
OB2 12 \citep{mcc98a, geb99}, WR 118 and HD 183143 \citep{mcc02},
and StRS 217 and W51 IRS 2 (Geballe et al. 2005, in preparation). If
we assume same velocity dispersion as the observed R(1, 1)$^l$
lines, those observations set the upper limit on the equivalent
width of $\sim$ 1 $\times$ 10$^{-6}$ $\mu$m and the level column
density of $\sim$ 3 $\times$ 10$^{13}$ cm$^{-2}$. The upper limit
for temperature depends on individual clouds. For Cygnus OB2 12 for
example, where the level column density of the (1, 1) level is 2
$\times$ 10$^{13}$ cm$^{-2}$ \citep{mcc02}, this sets the upper
limit of $\sim$ 100 K. The hot and diffuse environments in which
metastable line arises seem to be unique to sightlines toward the
Galactic center and, we believe, to regions within the CMZ.

\section{Discussion}

The semi-quantitative argument for the existence of hot and diffuse gas in
the CMZ given in the previous section can be more rigorously quantified
for the sightline to GCS 3-2 by comparing the high quality data there with
the model calculation of H$_3^+$ thermalization given in Paper I.
According to the calculation, the population ratios $N$(3, 3)/$N$(1, 1)
and $N$(3, 3)/$N$(2, 2) give definitive information on the temperature and
density, respectively, of such clouds as shown qualitatively in the
previous section. Also, the excitation temperature determined from the
population ratio $N$(1, 0)/$N$(1, 1) = 2 exp ($-$32.9/$T_{ex}$) gives
independent (but less accurate) information on the temperature and
density.

\subsection{Temperature and Density of Clouds toward GCS 3-2}

For each cloud component toward GCS 3-2, the local standard of rest
velocity $v_{LSR}$, the $v_{LSR}$ range, the equivalent width
$W_\lambda = \int (\Delta I(\lambda)/I)d\lambda$, and the
corresponding H$_3^+$ column density in the lower level of
absorption $N$(H$_3^+$)$_{level}$ =
(3hc/8$\pi^3\lambda$)$W_\lambda$/$\mid\mu\mid^2$ are listed in Table
2. The wavelength $\lambda$ and the strength $\mid\mu\mid^2$ of the
$R$(1, 1)$^l$, $R$(3, 3)$^l$, and $R$(2, 2)$^l$ transitions in D$^2$
are also given. As noted earlier, the $R$(1, 1)$^l$ spectrum is
composed of sharp absorption lines due to H$_3^+$ in ``ordinary"
clouds coexisting with CO in the intervening spiral arms, and broad
absorptions consisting of features at high and intermediate negative
velocities and the trough at lower velocities, all of which are
thought to be in the CMZ. We have separated the narrow and broad
absorptions as shown in Fig.~4, using the close similarity between
the sharp portion of the H$_3^+$ spectrum and the $R$(1) CO spectrum
in Fig.~1. Measured values of $W_{\lambda}$ and $N$(H$_3^+$) are
listed separately in Table 2. The values in parentheses are for
sharp absorptions. Separating the broad spectrum into cloud
components is difficult and somewhat arbitrary. The absorption
trough discussed above may consist of several components, but we can
find no evidence in our data for a clear-cut separation. For the
$-$123 km s$^{-1}$ and $-$97 km s$^{-1}$ clouds where a
deconvolution into gaussian components is feasible, the results are
similar to those obtained by separating the absorption into
$v_{LSR}$ ranges.

Thus our separation of the hot and diffuse gas is into what we believe are
the three most basic components: (1) clouds with $v_{LSR}$ $\sim$ $-$100
km s$^{-1}$ (from $-$140 to $-$74 km s$^{-1}$ in Table 2) which are likely
in the EMR; (2) the cloud with $v_{LSR}$ $\sim$ $-$50 km s$^{-1}$ (from
$-$74 to $-$40 km s$^{-1}$) which is the same velocity as the ``3 kpc
arm"; and (3) clouds with $v_{LSR}$ near 0 km s$^{-1}$ (from $-$ 40 to +
32 km s$^{-1}$), the aformentoned absorption trough. Observed values of
$N$(H$_3^+$)$_{level}$ for the (1, 1), (3, 3), (2, 2), and (1, 0)
rotational levels in each of these components are listed in Table 3. Those
are the only levels that are significantly populated apart from probable
small contributions from high metastable rotational levels ($J$, $J$) with
$J$ = 4, 5, and 6. The values for the (1, 0) level are from our Subaru
observations of the $Q$(1, 0) spectrum \citep{got02}. The separation of
broad features from the sharp lines in this spectrum is not as clear as in
the $R$(1, 1)$^l$ spectrum because of the lower spectral resolution and
has high uncertainties for the $-$50 km s$^{-1}$ and 0 km s$^{-1}$
clouds.

To determine the temperatures and densities of the $-$100, $-$50,
and 0 km s$^{-1}$ clouds, we use Fig. 5 of Paper I, which gives $T$
and $n$ as a function of the observed population ratios $N$(3,
3)/$N$(1, 1) and $N$(3, 3)/$N$(2, 2). The figure is reproduced in
Fig. 5 where the results of observation and analysis for the $-$ 100
km s$^{-1}$ cloud is shown as shaded area. The values of $N$(3,
3)/$N$(1, 1) and $N$(3, 3)/$N$(2, 2) used are 0.63 $\pm$ 0.15 and
$\geq 6 \pm 2$, respectively. Similar procedures have been used for
other clouds. The results of temperature and densities are shown in
the last columns of Table 3. It is seen that all clouds have
densities $\leq$ 200 cm$^{-3}$. The temperatures of the $-$100 km
s$^{-1}$ and the $-$50 km s$^{-1}$ clouds are high, $\geq$ 250 K.
The temperature of the 0 km s$^{-1}$ clouds is lower, although still
higher than ordinary diffuse clouds in the Galactic disk which have
been determined from the observed values of $N$(1, 0)/$N$(1, 1) to
be 25 - 50 K \citep{mcc02}. The excitation temperature calculated
from $N$(1, 0)/$N$(1, 1) provides an independent (but less direct)
measure of cloud temperature (see Fig.~6 of Paper I). It gives
$\sim$ 300 - 400 K for the $-$100 km s$^{-1}$ clouds and $\sim$ 200
- 250 K for the $-$50 km s$^{-1}$ cloud. In view of the larger
uncertainties in this method of determination, both in the
measurement and in the model calculations, the results are in good
agreement with those determined from $N$(3, 3)/$N$(1, 1).

\subsection{Accuracy}

The uncertainties in temperatures and densities listed in Table 3 are
standard deviations, but larger systematic errors are expected because of
inaccuracies in the model calculation of Paper I. Four general sources of
systematic error are known: (1) the steady state approximation, (2) the
theoretical rates of spontaneous emissions, (3) the assumed rates of
collision-induced rotational transitions given in Eq.(2) of Paper I, and
(4) neglect of hydrogen atoms as collision partners. As noted in Paper I,
we believe that errors introduced from (1) and (2) are at most 10 \% and
are minor. Assumptions (3) and (4) each introduce error mainly into the
determination of number density.

Assumption (3) has three sources of error: (a) the assumption of
completely random selection rules, (b) the formula given in Eq.(2),
and (c) use of the Langevin rate as the total collision rate. Out of
these three, (c) is most liable to introduce large overall errors.
It is based on the microwave pressure broadening measurement of the
HCO$^+$ - H$_2$ collision by \citet{and80} and the extensive
experimental and theoretical studies of the collision by the group
of de Lucia and Herbst \citep{pea95, lia96, oes01}. The latter
papers show that the collision rate constant is close to the
Langevin rate constant of 1.5 $\times$ 10$^{-9}$ cm$^3$ s$^{-1}$ for
the temperature range 10 - 77 K both experimentally and
theoretically. How well this applies to collisions of H$_3^+$, whose
rotational energy separations are much larger than those of HCO$^+$,
remains to be seen. Since experimental measurement of H$_3^+$
pressure broadening is next to impossible, theoretical calculations
are needed. Better still, a theoretical calculation of
state-to-state transition probabilities induced by collisions will
make the model calculation more accurate since it also will
eliminate errors caused by assumptions (a) and (b). If the H$_3^+$ -
H$_2$ collision rate constant is much lower than the Langevin rate,
the number densities of the clouds determined above will be an
underestimate. We believe that this error is no more than a factor
of two, especially at the cloud temperatures of $\sim$250 K. Since a
collision rate constant for a specific transition ($J', K'$)
$\rightarrow$ $(J, K)$,  $k^{J'K'}_{JK}$ is proportional to the
overall rate constant $C$ (which we set to the Langevin rate) and is
always multiplied with the cloud number density $n$(H$_2$), a
variation of $C$ does not change Fig. 5 except the scale of the
number density changes inversely proportional to $C$.

A considerable fraction of the hydrogen in the region of hot and
diffuse gas may be atomic due to photodissociation of H$_2$,
although because of the high volume density of molecular clouds in
the CMZ even the intercloud hydrogen may exist mainly in molecular
form (\citet{bin94} quoted in \citet{mez96}). However, since the
polarizability of H is comparable to that of H$_2$ (0.67 \AA$^3$
versus 0.79 \AA$^3$), the Langevin rate constants for the H$_3^+$ -
H and H$_3^+$ - H$_2$ collisions are comparable (2.2 $\times$
10$^{-}9$ cm$^3$ s$^{-1}$ versus 1.9 $\times$ 10$^{-9}$ cm$^3$
s$^{-1}$). Therefore, as long as we interpret the cloud density $n$
as $n$(H$_2$) + $n$(H), the results of Paper I are a good
approximation. Because of the low polarizability of He (0.20
\AA$^3$) and the higher reduced mass, the Langevin rate constant of
the H$_3^+$ - He collision is considerably lower (0.8 $\times$
10$^{-9}$ cm$^3$ s$^{-1}$) and its effect is negligible. Thus, we
believe that our derivation of low cloud densities, on the order of
100 cm$^{-3}$, is secure.

Our derived high temperatures are less affected by the model calculation
since they are based on the high energy of the (3, 3) metastable
level, unless there is some special pumping mechanism to influence the
non-thermal distribution. We have not been able to find such a mechanism.
Unlike in H$_2$ or C$_2$, optical pumping is inconceivable since H$_3^+$
does not have stable electronic excited state. Infrared pumping is very
unlikely since the H$_3^+$ vibrational transitions are isolated in the
middle of the L window from transitions of other (abundant) molecules.
Collisional pumping by the $J$ = 2 $\rightarrow$ 0 transition of H$_2$ at
510 K may discriminate between the (3, 3) and (2, 2) levels to some
extent, but is unlikely to cause such a drastically non-thermal population
as is observed.

\subsection{Total H$_3^+$ Column Densities}

According to the model calculation of Paper I, the (1, 1), (1, 0) and (3,
3) rotational levels contain most of the H$_3^+$ population, with only a
very small fraction of the H$_3^+$ in the (2, 2) level at the temperatures
and densities listed in Table 3. The higher metastable levels (4, 4), (5,
5), and (6, 6) contain a small but non-negligible fraction of the
population. From Fig. 5 of Paper I we estimate the total column densities
in those three metastable levels to be 1.4 $\times$ 10$^{14}$ cm$^{-2}$,
0.4 $\times$ 10$^{14}$ cm$^{-2}$, and 0.1 $\times$ 10$^{14}$ cm$^{-2}$,
for the $-$100 km s$^{-1}$, $-$50 km s$^{-1}$, and 0 km s$^{-1}$ clouds,
respectively.

The total H$_3^+$ column density toward GCS 3-2 including the cold clouds
in spiral arms is obtained by summing the total column densities of the
(1, 1), (3, 3) and (2, 2) levels given at the bottom of Table 2, the total
H$_3^+$ column density for the (1, 0) level of (12.1 $\pm$ 1.5) $\times$
10$^{14}$ cm$^{-2}$ determined from the $Q$(1, 0) line (see Table 3 of
\citet{got02}), and the column density of the high metastable levels
estimated above, (1.9 $\pm$ 0.5) $\times$ 10$^{14}$ cm$^{-2}$. The result
is (4.3 $\pm$ 0.3) $\times$ 10$^{15}$ cm$^{-2}$ in agreement with our
previous value of 4.6 $\times$ 10$^{15}$ cm$^{-2}$ \citep{got02}. The
total H$_3^+$ column density in two lowest rotational levels (1, 1) and
(1, 0), (3.3 $\pm$ 0.3) $\times$ 10$^{15}$ cm$^{-2}$, is also in
approximate agreement with our earlier value of 2.8 $\times$ 10$^{15}$
cm$^{-2}$ \citep{geb99}. Out of the total column density of (4.3 $\pm$
0.3) $\times$ 10$^{15}$ cm$^{-2}$, 1.2 $\times$ 10$^{15}$ cm$^{-2}$ is due
to H$_3^+$ in the cold clouds, mostly in the intervening spiral arms
while, 3.1 $\times$ 10$^{15}$ cm$^{-2}$ is in the hotter clouds in the
CMZ.

The visual extinction toward GCS 3-2 is calculated to be $A_V$
$\sim$ 30 mag from Fig.~20 of \citet{cot00}. How much of this
extinction is due to dense ($n \ge$ 10$^{3}$ cm$^{-3}$) clouds and
how much to diffuse clouds is not known.  If the ratio is like that
toward the infrared sources near Sgr A*, i.e., 1 : 2 \citep{whi97},
the visible extinction of A$_V$ $\sim$ 10 mag inferred for dense
clouds implies only a marginally detectable H$_3^+$ column density
if that extinction arises in the kind of dense cloud cores in which
H$_3^+$ has been found previously. This is because, as pointed out
in Section 1, the H$_3^+$ column density per unit extinction in
dense cores is 10 times smaller than in diffuse clouds. We note
however, that dense clouds encountered in spiral arms along the line
of sight are likely to be less dense than the ones in which H$_3^+$
has been found to date. As the number density of H$_3^+$ is constant
in dark clouds \citep{geb96, mcc99}, such lower density ``dense
clouds" may contain considerably more H$_3^+$ per unit extinction.
Thus we cannot constrain the cloud environment of the cold H$_3^+$
outside of the CMZ.

\subsection{Other Properties of the Clouds in the CMZ}

Observations toward many more sightlines covering a wider regions of
the CMZ are needed in order to elucidate the natures of the hot and
diffuse clouds in the CMZ besides their temperatures and densities.
Nevertheless, we can speculate on some physical and chemical
properties of the clouds based on the data obtained so far and on
the understanding of H$_3^+$ in the diffuse interstellar medium
gained from our previous studies.

One of the salient properties of H$_3^+$ as an astrophysical probe
is the simplicity and generality of its chemistry. This allows us to
express its number density in diffuse clouds in terms of the number
densities of H$_2$ and electron as $n$(H$_3^+$) =
($\zeta$/$k_e$)[$n$(H$_2$)/$n$(e)], where $\zeta$ is the ionization
rate of H$_2$ and $k_e$ is the rate constant for dissociative
recombination of H$_3^+$ with electrons \citep{mcc98a, mcc98b,
geb99, mcc02}. This formula demonstrates the remarkable property of
$n$(H$_3^+$) in diffuse clouds, that it is independent of the cloud
density as long as the ratio $n$(H$_2$)/$n$(e) is approximately
constant in the cloud. The H$_3^+$ column density can thus be
expressed as $N$(H$_3^+$) = $n$(H$_3^+$)$L$ in terms of the column
length $L$ in good approximation and we obtain

$\zeta$$L$ = $k_e$$N$(H$_3^+$)[$n$(e)/$n$(H$_2$)] =
2$k$$_e$$N$(H$_3^+$)($n_C$/$n_H$)$_{SV}$$R_X$/$f$,

\noindent where it is assumed that carbon is mostly in atomic form
(consistent with the absence of broad CO absorption in Fig.~1) and
is nearly completely singly ionized due to its low ionization
potential \citep{wel01, lis03}. In the above equation $f$ is the
fraction of hydrogen atoms in molecular form, $f$ =
2$n$(H$_2$)/[2$n$(H$_2$) + $n$(H)] $\leq$ 1. The gas phase carbon to
hydrogen ratio after depletion in the solar vicinity,
($n_C$/$n_H$)$_{SV}$ = 1.6 $\times$ 10$^{-4}$ \citep{sof04}, is
multiplied by $R_X$ = ($n_C$/$n_H$)$_{GC}$/($n_C$/$n_H$)$_{SV}$
which takes into account the higher carbon to hydrogen ratio near
the Galactic center than in the solar vicinity. The value of $R_X$
has been given as 3 $-$ 10 by the COBE Diffuse Infrared Background
Experiment by \citet{sod95} and as ``at least 3" by \citet{ari96}
\citep{mez96}.

\subsubsection{The $-$100 km s$^{-1}$ clouds}

The hot and diffuse clouds with high velocities of $-$123 km
s$^{-1}$ and $-$97 km s$^{-1}$, most likely associated with the EMR,
have more than half of H$_3^+$ in the CMZ and provide the most
definitive information since their absorption features are least
contaminated by that of cold H$_3^+$ in intervening spiral arms.
Using the total column density of (15.7 $\pm$ 1.7) $\times$
10$^{14}$ cm$^{-2}$ listed in Table 3 and $k_e$ = 7.3 $\times
10^{-8}$ cm$^3$ s$^{-1}$ for $T$ = 270 K calculated from Eq. (7) of
\citet{mcc04}, we obtain

             $\zeta$$L$ = 3.7 $\times$ $10^4$ $R_X$/$f$ cm s$^{-1}$.

For the canonical value of the interstellar cosmic ray ionization
rate, $\zeta = 3 \times 10^{-17}$ s$^{-1}$ (see for example
\citet{van86} and \citet{lee96}), this formula gives a huge and
clearly unreasonable value of $L$ $>$ 1 kpc, for $R_X$ $\geq$ 3,
even if $f$ = 1. This situation is the same as in the diffuse clouds
in the Galactic disk \citep{mcc98a, geb99, mcc02}. \citet{mcc03}
deduced that $\zeta$ = 1.2 $\times$ 10$^{-15}$ s$^{-1}$, 40 times
higher than the canonical value of $\zeta$, to acccount for the
H$_3^+$ column density measured in the classic visible sightline
toward $\zeta$ Per. Such a high ionization rate is far beyond the
limit set from observed abundances of HD and OH by \citet{har78a,
har78b}. However, recent original work by \citet{lis03} has shown
that values of $\zeta$ beyond the canonical value are favored in a
wide range of circumstances if the effect of neutralization of
atomic ions on grains is properly taken into account. Subsequently
\citet{lep04} reported a chemical model calculation using a lower
value of $\zeta$ = 2.5 $\times$ 10$^{-16}$ s$^{-1}$ and explained
observed abundances of many molecules toward $\zeta$ Per including
HD and OH. As far as H$_3^+$ is concerned, however, their result is
nearly identical to that of \citet{mcc03} since they assumed 1.9
times longer path (4 pc versus 2.1 pc), 4.8 times lower $\zeta$, and
obtained an H$_3^+$ column density which is lower than the observed
by a factor of 2.8 (2.9 $\times$ 10$^{13}$ cm$^{-2}$ versus 8.0
$\times$ 10$^{13}$ cm$^{-2}$). We believe that observed abundance of
H$_3^+$ gives more direct information on $\zeta$. \citet{mcc03} used
$n$(e)/$n$(H$_2$) = 3.8 $\times$ 10$^{-4}$ obtained directly from
the observed C$^+$ and H$_2$ column densities, which corresponds to
$f$ $\sim$ 0.85. The values of $f$ for clouds in the CMZ are not
known. The absence of H~I absorption in the spectrum of the Arched
Filament complex \citep{lan04} and the small absorption in the
Sickle near the sightline of the quintuplet (C. C. Lang private
communication) suggest that $f$ is between 1 and 1/2. Lower values
give longer pathlengths.

Based on the $^{13}$CO observations by \citet{bal87, bal88},
\citet{sof95} estimated the thickness of the EMS to be $\sim$ 15 pc.
If we use the values of $f$ and $\zeta$ by \citet{mcc03}, we obtain
$L$ $\sim$ 12$R_X$ pc for the $-$100 km s$^{-1}$ clouds.  If the
value of $R_X$ is 3 - 10 \citep{sod95}, the pathlength is 36 - 120
pc much higher than Sofue's estimate. This suggests that the
ionization rate $\zeta$ in the CMZ is even higher than the value by
\citet{mcc03} for the diffuse interstellar medium in the Galactic
disk. In order to make the pathlength on the order of 20 pc, the
value of the ionization rate must be (2 - 7) $\times$ 10$^{-15}$
s$^{-1}$. Such a high ionization rate, however, will introduce other
consequences discussed by \citet{lis03}. In particular ionization of
H increases the electron density and decreases the H$_3^+$ number
density. It may well be that the simple linear relation between
$\zeta$ and $N$(H$_3^+$) used in this section needs to be modified
for the Galactic center.

\subsubsection{The $-$50 km s$^{-1}$ cloud}

Although we believe that this cloud lies within the CMZ, the precise
radial location is uncertain, and whether or not the agreement of
its velocity and that of the 3 kpc arm is accidental remains to be
seen. The 3 kpc spiral arm merges into the Galactic bar and reaches
the CMZ (Fig.~4 of \citet{mez96}) but how the velocity varies with
radius is not known. If the cloud is in the CMZ, its rather high
negative velocity suggests an association with the EMR. Observations
of additional Galactic center infrared sources may provide a clearer
picture of this cloud.

The $-$50 km s$^{-1}$ cloud has a slightly lower temperature and
higher density than the $-$100 km s$^{-1}$ clouds. The observed
total H$_3^+$ column density of the cloud, (6.6 $\pm$ 1.3) $\times$
10$^{14}$ cm$^{-2}$ and the recombination rate constant of $k_e$ =
7.6 $\times$ 10$^{-8}$ cm$^3$ s$^{-1}$ at 250 K gives

    $\zeta$$L$ = 1.6 $\times$ $10^4$ $R_X$/$f$ cm s$^{-1}$,

\noindent indicating that the column length of this cloud is shorter than
the $-$100 km s$^{-1}$ cloud by a factor of approximately 2.3. The
chemical condition of this cloud is likely similar to that of the $-$100
km s$^{-1}$ clouds. The ionization rate in it must be comparable to the
value discussed above for the $-$100 km s$^{-1}$ clouds.

\subsubsection{The 0 km s$^{-1}$ clouds}

This is probably a complex of many clouds in the CMZ located at various
distances from the Galactic nucleus with lower temperature and higher
density than the $-$100 km s$^{-1}$ and $-$50 km s$^{-1}$ clouds. Some of
the clouds may be associated with the Quintuplet cluster. The temperature
listed in Table 3 is the average value for the entire velocity range, and
thus some of the 0 km s$^{-1}$ clouds may have temperatures outside the
uncertainties in Table 3. In addition, some may have densities higher than
200 cm$^{-3}$. Nevertheless, the average value indicates that most clouds
are diffuse and their temperatures are higher than the usual diffuse
clouds in the Galactic disk. Observations of more sightlines toward
infrared stars in the CMZ may enable us to separate some of these clouds.

The observed total H$_3^+$ column density of these clouds, (8.4
$\pm$ 1.6) $\times$ 10$^{14}$ cm$^{-2}$, and the recombination rate
of 1.09 $\times 10^{-7}$ cm$^3$ s$^{-1}$ at 130 K gives

    $\zeta$$L$ = 2.9 $\times$ $10^4$ $R_X$/$f$ cm s$^{-1}$,

\noindent indicating that the total path length of those clouds is
shorter than the $-$100 km s$^{-1}$ cloud by a factor of $\sim$ 1.3
if the same value of $\zeta$ and $R_X$/$f$ are assumed. Again the
value of $L$ is too high unless we assume $\zeta$ to be higher than
the value in diffuse clouds in the Galactic disk. The Quintuplet
cluster contains numerous hot stars with luminosities of $\sim$
10$^5$ $L_{\odot}$ \citep{fig99} implying a high rate of
photo-ionization in the nearby interstellar medium. Moreover the
strong X-rays in the CMZ will photoionize the gas. Also its location
near the Radio Arc suggests high MHD activity. Thus it would not be
surprising if the value of $\zeta$ there were much higher than in
diffuse clouds in the Galactic disk.

Finally, note that discussions of the cloud dimension $L$ and the
ionization rate $\zeta$ in this section 4.4 are based on the
dissociative recombination rate constant $k_e$ reported by
\citet{mcc03, mcc04}. While this value has also been supported
theoretically by \citet{kok03a, kok03b}, there still is a
possibility that the value is not final. Since laboratory
experiments cannot completely duplicate interstellar conditions,
further theoretical confirmation is highly desirable.

\subsection{The Dense Cloud Filling Factor in the CMZ}

The foregoing analysis shows that almost all of the H$_3^+$ observed
in the CMZ is in diffuse clouds with densities on the order of 100
cm$^{-3}$. This conclusion is not surprising since H$_3^+$ is a more
sensitive probe for diffuse clouds than for dense clouds (in terms
of column density of H$_3^+$ per unit extinction; see Section 1),
and that the extinction toward the GC is mostly due to diffuse
clouds. It shows, however, that the amount of dense cloud material
in the CMZ is much less than previously thought.  If indeed the CMZ
contains high density gas ($\geq$ 10$^4$ cm$^{-3}$) with a high
volume filling factor ($\geq$ 0.1) as mentioned in \citet{mor96}, we
would have easily detected H$_3^+$. Such a filling factor would
produce a mean pathlength of 20 pc of dense gas along the line of
sight to sources in the CMZ. Dense clouds in the Galactic disk
surrounding AFGL 2136 and W 33A with a pathlengths of 1.3 pc and 1.7
pc showed H$_3^+$ column densities of 3.8 $\times$ 10$^{14}$
cm$^{-2}$ and 5.2 $\times$ 10$^{14}$ cm$^{-2}$, respectively
\citep{mcc99}. Thus a very strong absorption by H$_3^+$ in dense CMZ
clouds would be expected, even if the velocity components are broad
and the metallicity is high.

Moreover, the above pathlength through dense molecular gas implies
an H$_2$ column density of 6 $\times$ 10$^{23}$ cm$^{-2}$, which
would be easy to detect through its infrared absorption spectrum
\citep{lac94} even if the cloud has a high velocity dispersion. Such
attempt has been unsuccessful (Usuda \& Goto, private
communication). \citet{rod01} reported ``a few 10$^{22}$ cm$^{-2}$"
toward 16 sources within 500 pc of the Galactic center.

While it is possible that the sightline toward GCS 3-2 and the other seven
stars studied in this paper happen to have low density as statistical
fluctuations, it is more likely that the combination of high density
($\geq$ 10$^4$ cm$^{-3}$) and large volume filling factor of ($\geq$ 0.1)
is a considerable overestimate. That combination gives a mass
of the gas in the CMZ which is much higher than 5 - 10 $\times$ 10$^7$
$M_{\bigodot}$ estimated from radio observations of molecules (see many
references quoted in \citet{mor96}). More recent papers of \citet{oka98b}
and \citet{tsu99}, using CO $J$ = 2 - 1 and CS $J$ = 1 - 0 lines have
reported mass values of 2 $\times$ 10$^7$ $M_{\bigodot}$ (as a lower
limit) and 3 - 8 $\times$ 10$^7$ $M_{\bigodot}$, respectively. Also the
large dense cloud path length of 20 pc would give a visual extinction
at least an order of magnitude higher than $A_V$ = 25 - 40
reported by \citet{cot00} for four large regions within 40 pc of the
Galactic center.

Out of many molecular emissions used to study the CMZ, the 1 - 0
emission line of CO and inversion spectrum of NH$_3$ cannot provide
evidence for dense clouds by themselves since their critical
densities are lower than 10$^4$ cm$^{-3}$ by more than an order of
magnitude. The CO 2 - 1 \citep{oka98b} and higher \citep{har85}
emission lines and the 1 - 0 emission lines of CS \citep{tsu99,
saw01}, HCN and HCO$^+$ \citep{lin81, wri01} do provide definitive
evidence for high density clouds. Also, high densities in the CMZ
have been reported from radio absorption of HCN and CS using model
calculations \citep{gre92, gre95}. However the filling factor of
these clouds is probably much lower than 0.1, more likely 0.01 or
even less. (In this regard the observation of broad
\emph{absorption} by HCO$^+$ by \citet{lin81} is noteworthy). The
fraction of $\sim$ 10 \% quoted as the ratio of gas in the CMZ to
the total gas content of the Galaxy (see for example \citet{mez96}
and \citet{dah98}) must also be a gross overestimate.

\section{Summary}

We have observed three absorption lines of H$_3^+$ towards several
infrared sources in the Galactic center, indicating the presence of
extensive hot and diffuse clouds with high velocity dispersions in
the inner 200 pc Central Molecular Zone of the Galaxy. The observed
high column densities of H$_3^+$ in the ($J$, $K$) = (3, 3)
metastable rotational level, which is 361 K above the (1, 1) ground
level, provides definitive evidence of the high temperature, and the
absence or small quantity in the (2, 2) level, which is 210 K lower
than the (3, 3) level, provides evidence for the low density of
these clouds. This remarkable non-thermal rotational distribution is
caused by the metastability of the (3, 3) level and the relatively
fast spontaneous emission from (2, 2) to (1, 1). The hot and diffuse
gas toward the brightest infrared source GCS 3-2 in the Quintuplet
Cluster has been quantitatively studied. We estimate that
three-fourths of the total H$_3^+$ column density of 4.3 $\times$
10$^{15}$ cm$^{-2}$ observed towards this source arises in the CMZ,
with half of this fraction associated with the 180 pc radius
Expanding Molecular Ring, and the remainder in lower velocity
clouds. These clouds have a range of temperatures and densities,
with all of the temperatures considerably higher than diffuse clouds
in the Galactic disk where the (3, 3) level has not been detected.
The remaining one-fourth of the observed H$_3^+$ is located outside
the CMZ, largely in spiral arms along the line of sight to the
nucleus.

In order to make the aggregate line of sight pathlength of the
diffuse CMZ gas a reasonable value, it is necessary to assume a very
high ionization rate $\zeta$ = (2 - 7) $\times$ 10$^{-15}$ s$^{-1}$.
The high metallicity in the CMZ is only partly responsible for the
need for the large increase of $\zeta$ over its value elsewhere in
the Galaxy. The high value is no doubt intimately related to the
exceptionally energetic environment of the Galactic center.

Almost all of H$_3^+$ in the CMZ exists in low density clouds. The filling
factor of high density clouds in the CMZ, $f$ $\geq$ 0.1, that has been
reported previously, must be a considerable overestimate.

Questions abound on the hot and diffuse clouds in the CMZ. How are they
related to dense clouds studied by neutral molecules and the H~II regions
studied by hydrogen recombination lines? What is their relation to the
sources of X-ray emission and to the strong magnetohydrodynamic phenomena
in the nucleus? What is their heating mechanism? How is the pressure
balanced? Additional observations of H$_3^+$ and CO toward other infrared
stars in the CMZ including those at larger distances from the Galactic
nuclei may help provide answers to some of these questions.

\acknowledgments

We thank the staffs of Subaru, Gemini, and the Joint Astronomy
Centre for their support of these observations. We are grateful to
T. Nagata for providing us information on the suitability of the NHS
stars for the H$_3^+$ observations, and to C. Lang for giving us the
H~I radio spectrum of the Sickle prior to publication. We thank C.
Morong for his help in drawing Fig. 5. We also are grateful to
Harvey Liszt and Tomoharu Oka for critical reading of this paper and
to H. Liszt, M. Morris, T. Nagata, G. Novak, T. Oka, M. Tsuboi, T.
L. Wilson, and F. Yusef-Zadeh for helpful discussions on the
Galactic center. T. O. acknowledges the NSF grant PHY-0354200. T. R.
G.'s research is supported by the Gemini Observatory, which is
operated by the Association of Universities for Research in
Astronomy, Inc., on behalf of the international Gemini partnership
of Argentina, Australia, Brazil, Canada, Chile, the United Kingdom
and the United States of America. M. G. is supported by a Japan
Society for Promotion of Science fellowship. B. J. M. has been
supported by a Camille and Henry Dreyfus Foundation New Faculty
Award.

 \clearpage



\begin{figure}
\epsscale{.80} \plotone{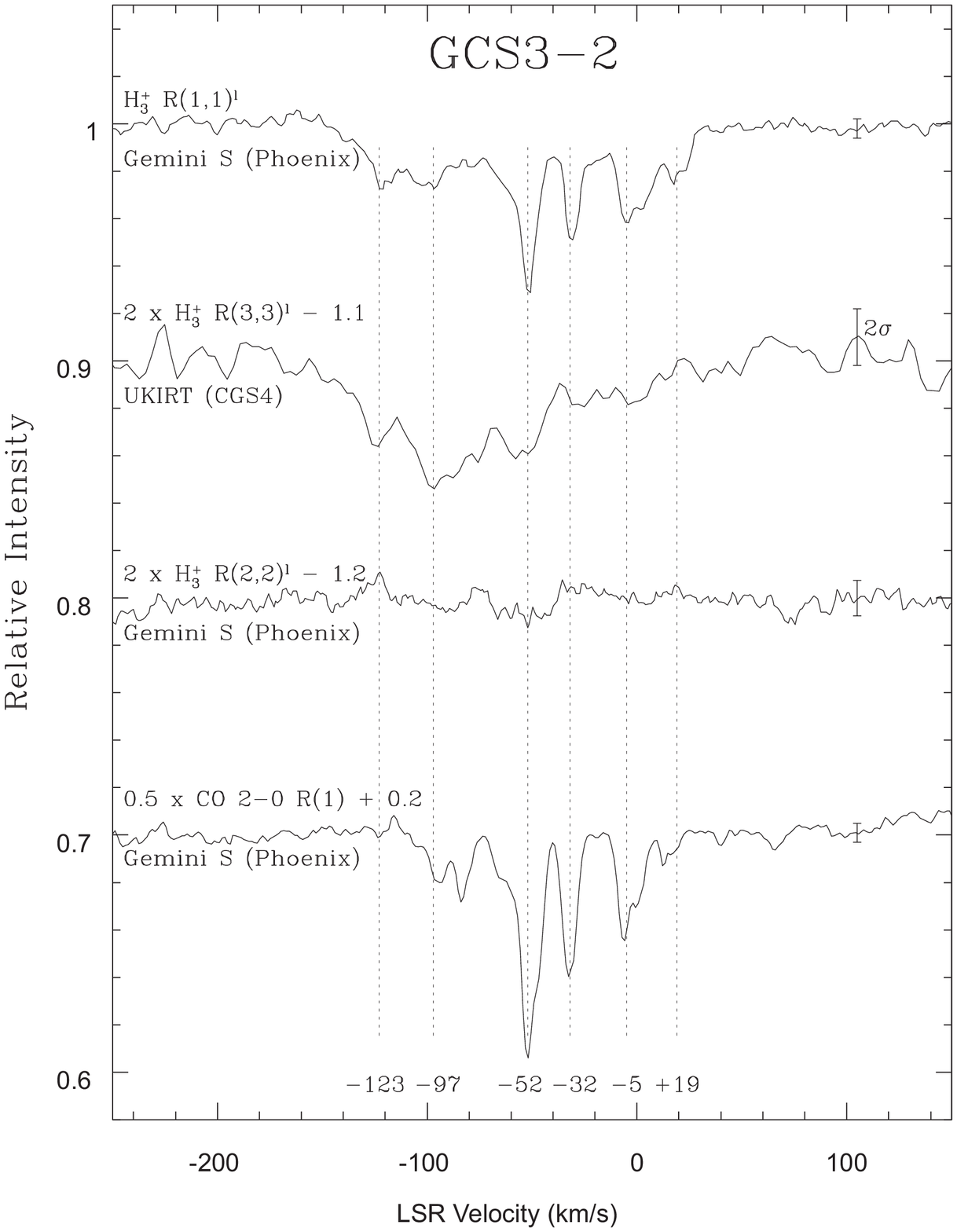} \caption{Observed H$_3^+$ (top
three) and CO spectra toward GCS 3-2. The three H$_3^+$ spectra are
(from the top) the $R$(1, 1)$^l$, $R$(3, 3)$^l$, and $R$(2, 2)$^l$
transitions, starting from the (1, 1) ground level, the (3, 3)
metastable level, and the (2, 2) unstable level, respectively. The
vertical scaling of the $R$(3, 3)$^l$ and $R$(2, 2)$^l$ spectra are
multiplied by a factor of 2 and that of the CO spectrum is divided
by 2 for clarity.} \label{fig1}
\end{figure}

\clearpage


\begin{figure}
\epsscale{.50} \plotone{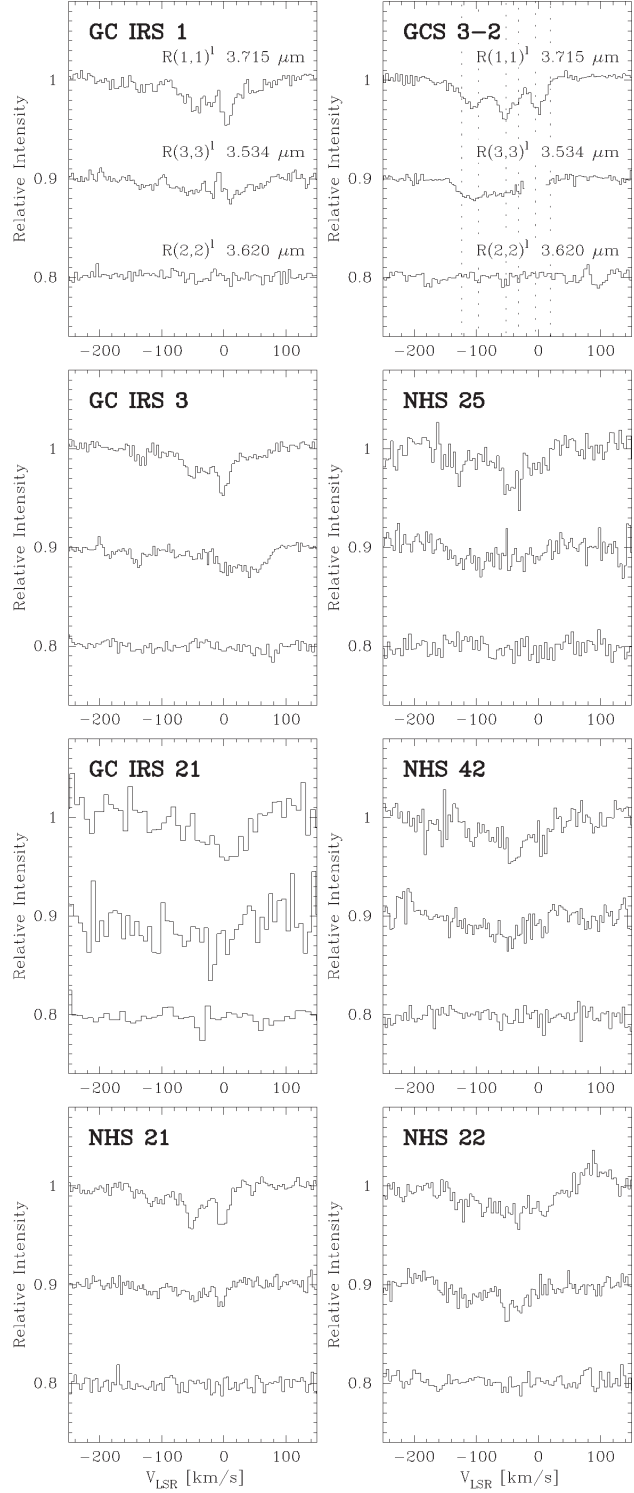}
\caption{The $R$(1, 1)$^l$, $R$(3, 3)$^l$, and $R$(2, 2)$^l$
spectral lines observed by the IRCS of Subaru toward bright infrared
sources in the CMZ. The noise can be estimated by point-to-point
fluctuations.\label{fig2}}

\end{figure}

\clearpage

\begin{figure}
\epsscale{1} \plotone{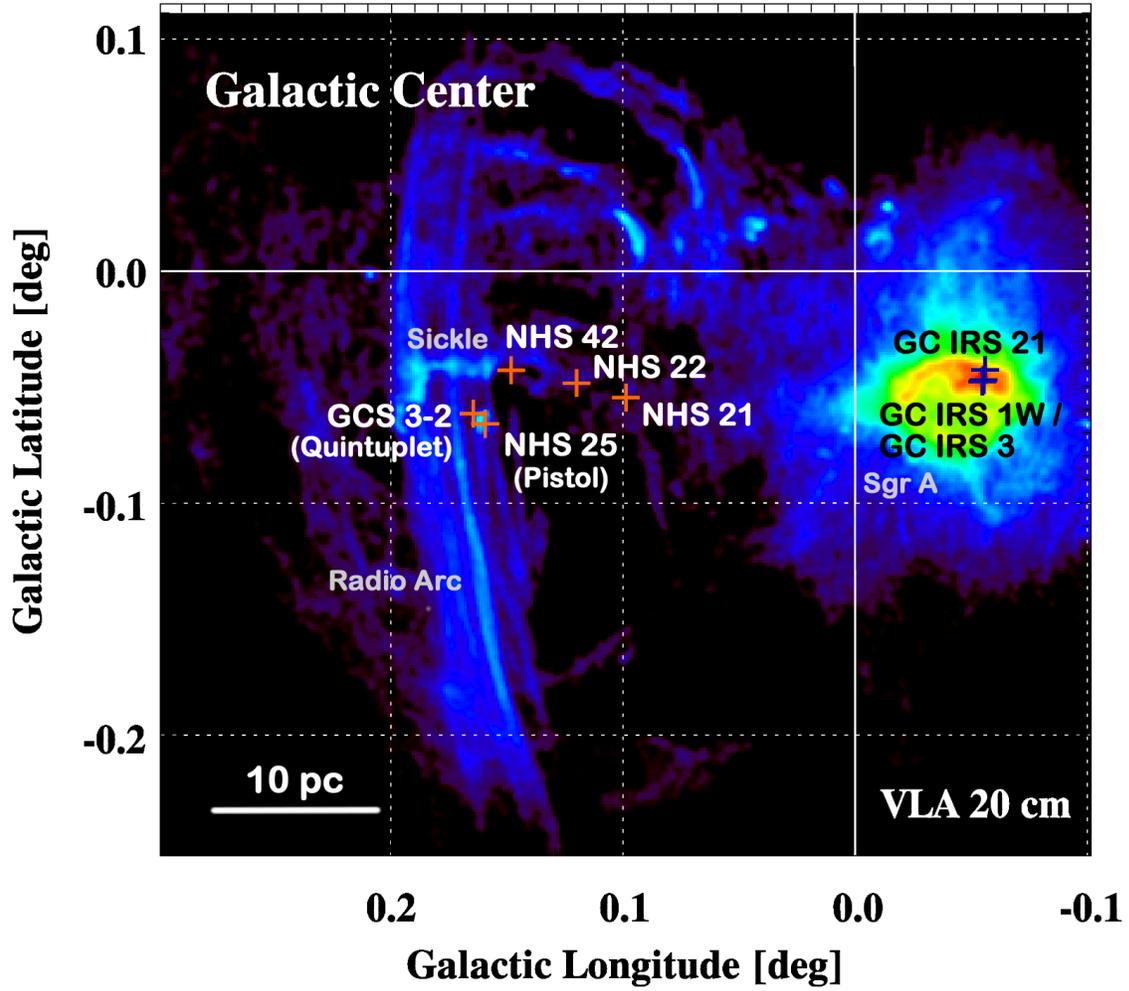} \caption{Location of infrared stars
toward which hot and diffuse clouds have been observed, plotted on
the VLA 20 cm image from \citet{yus87}. \label{fig3}}
\end{figure}

\clearpage

\begin{figure}
\epsscale{1} \plotone{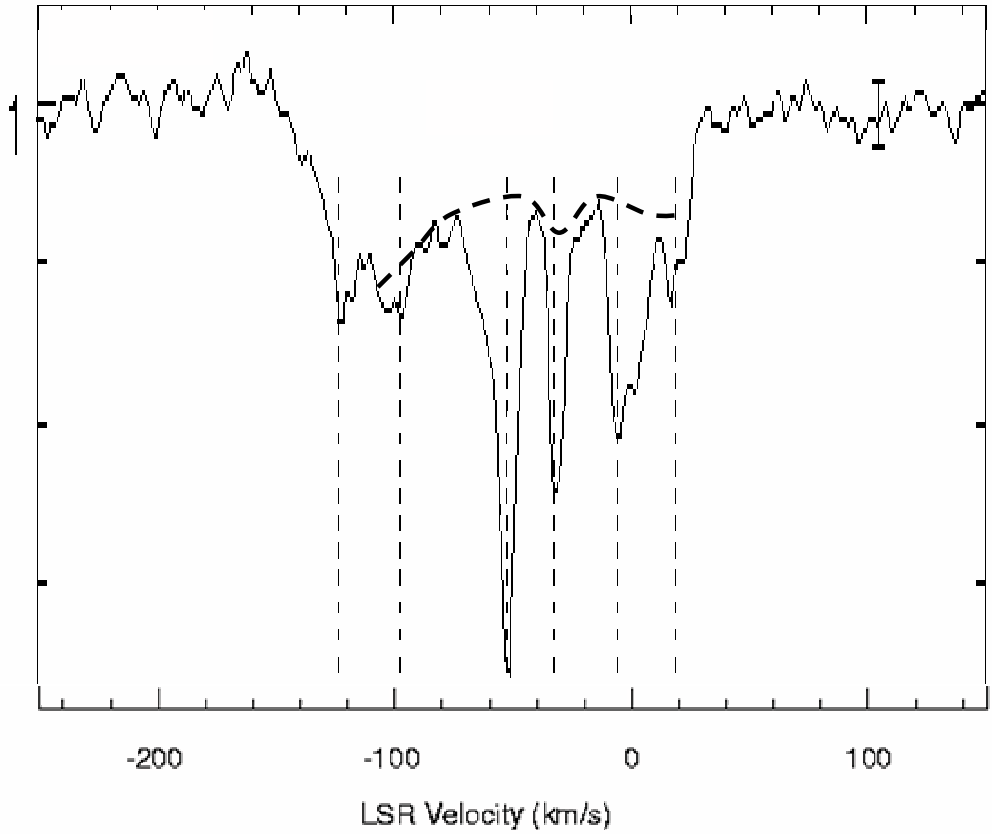} \caption{The H$_3^+$ $R$(1, 1)$^l$
spectrum toward GCS 3-2 (the top trace of Fig. 1) separated into
sharp components caused by cold H$_3^+$ in intervening spiral arms
and broad components arising in hot and diffuse clouds with high
velocity dispersion that are most likely in the CMZ. \label{fig4}}
\end{figure}

\clearpage

\begin{figure}
\epsscale{1} \plotone{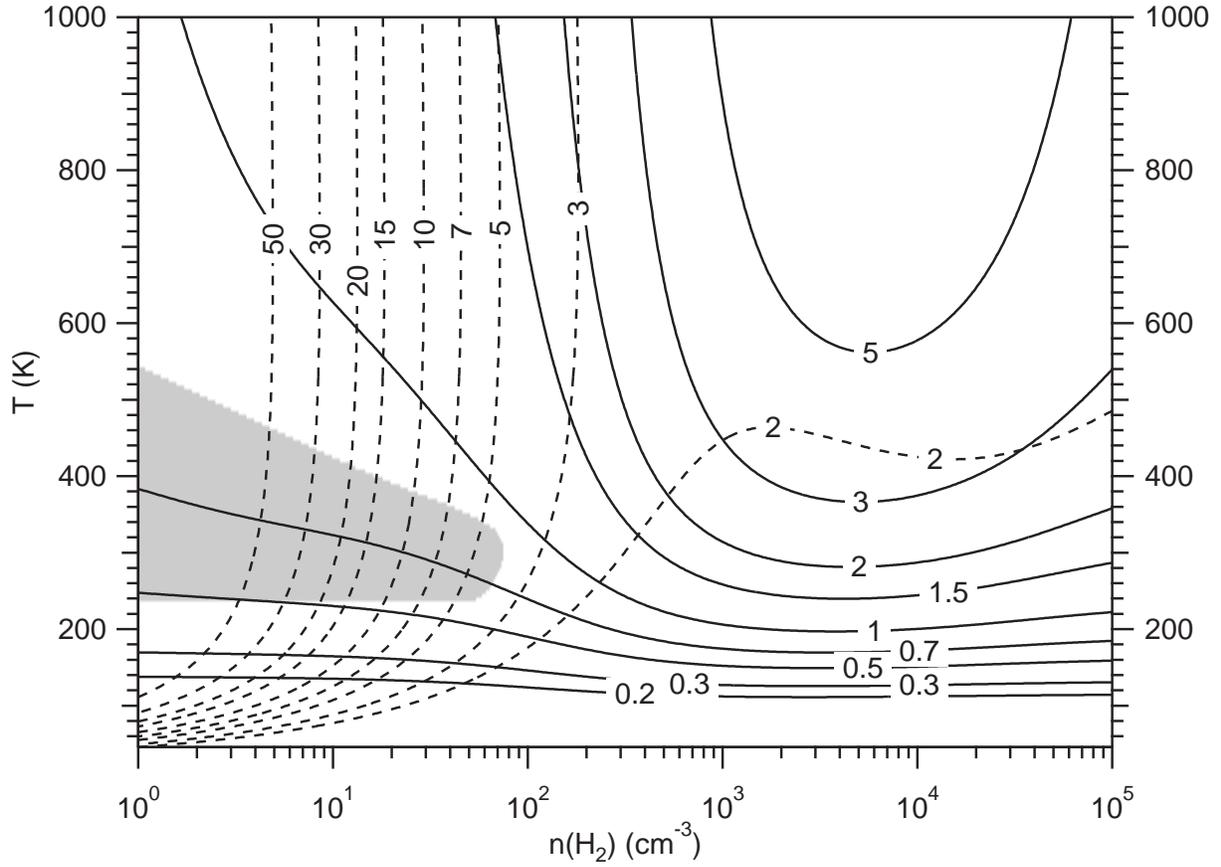}
\caption{A plot of the population ratio $N$(3, 3)/$N$(1, 1) (solid
lines) and $N$(3, 3)/$N$(2, 2) (broken lines) as a function of cloud
density $n$(H$_2$) and kinetic temperature $T$ given in Paper I. The
observed range for the $-$ 100 km s$^{-1}$ cloud is shown by the
shaded area.\label{fig5}}
\end{figure}







\clearpage

\begin{deluxetable}{llllllrlllc}

\tabletypesize{\scriptsize}

\rotate

\tablecaption{Log of observations} \tablewidth{0pt} \tablehead{
\colhead{UT date} & \colhead{Telescope} & \colhead{Instrument} &
\colhead{Object} & \colhead{RA(2000)} & \colhead{D(2000) } &
\colhead{$l$($^{\circ}$)} & \colhead{$b$($^{\circ}$)} &
\colhead{Spectrum} & \colhead{$\lambda(\mu$m)} &
\colhead{time(min)} } \startdata

2003 Jul 23 & Gemini S & Phoenix & GCS 3-2 &17 46 14.9& -28 49 43&
0.16 & -0.06& H$_3^+$ $R$(1, 1)$^l$
& 3.7155 & 20 \\
2003 Jul 24 & Gemini S & Phoenix & GCS 3-2 &&&&& H$_3^+$ $R$(2,
2)$^l$
& 3.6205 & 30  \\
2004 Apr 2 & Gemini S & Phoenix & GCS 3-2 &&&&& CO $R$(0)-$R$(3) &
2.342& 8 \\

2004 Jul 7 & Subaru & IRCS & GCS 3-2 &&&&& H$_3^+$ multiline &
--- &
45 \\
& Subaru & IRCS & GC IRS 3 & 17 45 39.6 & -29 00 24 & -0.06&
-0.04& H$_3^+$ multiline &
--- &
52 \\

2004 Jul 8 & Subaru & IRCS & GC IRS 3 &&&&& H$_3^+$ multiline &
--- &
40 \\
& Subaru & IRCS & NHS 21 & 17 46 04.3 & -28 52 49 & 0.10 & -0.06 &
H$_3^+$ multiline &
--- &
100 \\

2004 Jul 27 & Subaru & IRCS & GC IRS 1W & 17 45 40.2& -29 00 27&
-0.06 & -0.05& H$_3^+$ multiline &
--- &
85 \\
& Subaru & IRCS & NHS 42 & 17 46 08.3& -28 49 55& 0.15& -0.04 &
H$_3^+$ multiline &
--- &
80 \\

2004 Jul 28 & Subaru & IRCS & GC IRS 21 & 17 45 40.2& -29 00 30&
-0.06& -0.05& H$_3^+$ multiline &
--- &
30 \\

2004 Aug 30 & UKIRT & CGS4 & GCS 3-2 &&&&& H$_3^+$ $R$(3, 3)$^l$ &
3.5337 & 8.2 \\

2004 Sep 1 & UKIRT & CGS4 & GCS 3-2 &&&&& H$_3^+$ $R$(2, 2)$^l$ &
3.6205 & 45 \\

2004 Sep 2 & Subaru & IRCS & GC IRS 21 &&&&& H$_3^+$ multiline &
--- &
40 \\

2004 Sep 4 & Subaru & IRCS & NHS 22 & 17 46 05.6& -28 51 32& 0.12&
-0.05& H$_3^+$ multiline &
--- &
40 \\
& Subaru & IRCS & NHS 25 & 17 46 15.3 & -28 50 04& 0.16& -0.07&
H$_3^+$ multiline &
--- &
40 \\

2004 Sep 25 & Subaru & IRCS & GC IRS 3 &&&&& H$_3^+$ multiline &
--- &
50 \\

2004 Sep 26 & Subaru & IRCS & GC IRS 3 &&&&& H$_3^+$ multiline &
--- &
32 \\

\enddata

\end{deluxetable}


\clearpage



\begin{deluxetable}{ccrccrcc}

\tabletypesize{\scriptsize}

\rotate

\tablecaption{Observed velocities, equivalent widths, and
H$_3^+$ column densities of clouds toward GCS 3-2}
\tablewidth{0pt} \tablehead{ \colhead{$v_{LSR}$} &
\colhead{Range}& & \colhead{$W_\lambda$  } &&&
\colhead{$N$(H$_3^+$)$_{level}$} &  } \startdata

[km s$^{-1}$] & [km s$^{-1}$] & & [10$^{-5} \mu$m] & && [10$^{14}$
cm$^{-2}$]& \\

 &&  $R$(1, 1)$^{l,a}$& $R$(3, 3)$^l$ & $R$(2, 2)$^l$
&(1, 1)$^b$ & (3, 3) & (2, 2)\\

\tableline - 123& - 140 $\rightarrow$ - 113 & 0.56 $\pm$ 0.10&
0.32 $\pm$ 0.14
& $\leq$ 0.07  & 2.6 $\pm$ 0.5 & 1.1 $\pm$ 0.5 & $\leq$ 0.3   \\

 - 97& - 113 $\rightarrow$ - 74  & (0.04) 0.96 $\pm$ 0.14 & 0.93
 $\pm$ 0.20
& $\leq$ 0.10  & (0.2) 4.4 $\pm$ 0.6 & 3.3 $\pm$ 0.7 & $\leq$ 0.4   \\

\tableline - 52& - 74 $\rightarrow$ - 40  & (0.63) 0.57 $\pm$ 0.12
& 0.46 $\pm$ 0.18
& 0.12 $\pm$ 0.11 &  (2.9) 2.6 $\pm$ 0.5 & 1.6 $\pm$ 0.6 & 0.4 $\pm$ 0.4  \\

\tableline - 33& - 40 $\rightarrow$ - 26 & (0.30) 0.27 $\pm$ 0.05
& 0.08 $\pm$ 0.07
& $\leq$ 0.05  & (1.4) 1.2 $\pm$ 0.2 & 0.3 $\pm$ 0.3 & $\leq$ 0.2   \\

& - 26 $\rightarrow$ - 13 & 0.23 $\pm$ 0.05 & 0.08 $\pm$ 0.07
& $\leq$ 0.05  & 1.1 $\pm$ 0.2 & 0.3 $\pm$ 0.3 & $\leq$ 0.2   \\

 - 5 & - 13 $\rightarrow$ +12 & (0.48) 0.40 $\pm$ 0.09 & 0.10
 $\pm$ 0.13
& $\leq$ 0.05  & (2.2) 1.8 $\pm$ 0.4 & 0.4 $\pm$ 0.5 & $\leq$ 0.2   \\

+19 & +12 $\rightarrow$ +32 & (0.03) 0.17 $\pm$ 0.05 & $\leq$ 0.05
& $\leq$ 0.04 & (0.1) 0.8 $\pm$ 0.2 & $\leq$ 0.2 & $\leq$ 0.1   \\

\tableline & Total & 4.64 $\pm$ 0.24 & 1.97 $\pm$ 0.34 & 0.12 $\pm$
0.11 & 21.3 $\pm$ 1.1
 & 7.0 $\pm$ 1.2 & 0.4 $\pm$ 0.4 \\

 \tableline\tableline

 & Transition & $\lambda$ ($\mu$m)  & $\mid\mu\mid^{2}$ (D$^2$)  &&&& \\
 & $R$(1, 1)$^l$ & 3.7155 &  0.01407 &&&& \\
 & $R$(3, 3)$^l$ & 3.5227 &  0.01914 &&&& \\
 & $R$(2, 2)$^l$ & 3.6205 &  0.01772 &&&& \\

\enddata



\tablenotetext{a}{The equivalent widths of the $R$(1, 1) spectrum
are separated into those of sharp features in parentheses and
broad features. No sharp features exists in the $R$(3, 3)$^l$ and
$R$(2, 2)$^l$ spectra.} \tablenotetext{b}{The column densities of
the (1, 1) level are separated into those of sharp features in
parentheses and broad features.}

\end{deluxetable}

\clearpage

\begin{deluxetable}{ccccccccccc}

\tabletypesize{\scriptsize}

\rotate

\tablecaption{H$_3^+$ column densities, temperatures and densities
of high velocity dispersion clouds toward GCS 3-2}\tablewidth{0pt}
\tablehead{ \colhead{Clouds} & \colhead{$v_{LSR}$}&  \colhead{Range
} &&& \colhead{$N$(H$_3^+$)$_{level}$  } & \colhead{ } & &
&\colhead{$T$} &\colhead{$n$} } \startdata

 & [km s$^{-1}$] & [km s$^{-1}$] & & & [10$^{14}$cm$^{-2}$] & & & & [K] & [cm$^{-3}$] \\
 &  &  & (1, 1) & (3, 3) & (2, 2) & (1, 0)& HM$^a$ & Total && \\
\tableline ``$-$100 km s$^{-1}$" &  $-$123, $-$97 & $-$140
$\rightarrow$ $-$74 & 7.0 $\pm$ 0.8 & 4.4 $\pm$ 0.9
& $\leq$ 0.7 & 2.9 $\pm$ 1.0 & 1.4 $\pm$ 0.7 & 15.7 $\pm$ 1.7 & 270 $\pm$ 70 & $\leq$ 50\\

``$-$50 km s$^{-1}$" & $-$52 & $-$74 $\rightarrow$ $-$40 & 2.6
$\pm$ 0.5 & 1.6 $\pm$ 0.6
& 0.4 $\pm$ 0.4 & 1.6 $\pm$ 0.9 & 0.4 $\pm$ 0.2 & 6.6 $\pm$ 1.3 & 250 $\pm$ 100 & $\leq$ 100   \\

``0 km s$^{-1}$" &  $-$33, $-$5, +19 & $-$40 $\rightarrow$ +32
& 4.9 $\pm$ 0.5 & 1.0 $\pm$ 0.7
& $\leq$ 0.7  &  2.4 $\pm$ 1.3& 0.1 $\pm$ 0.1 & 8.4 $\pm$ 1.6 & 130 $\pm$ 100 & $\leq$ 200   \\

\tableline

Total & & & 14.5 $\pm$ 1.1 & 7.0 $\pm$ 1.3 & 0.4 $\pm$ 0.4 & 6.9
$\pm$ 1.9 & 1.9 $\pm$ 0.7 & 30.7 $\pm$ 2.7 & & \\
\enddata

\tablenotetext{a}{Sum of calculated H$_3^+$
column densities for high metastable levels (4, 4), (5, 5), and
(6, 6).}





\end{deluxetable}





\clearpage

\begin{thebibliography}{}

\bibitem[Adamson et al.(2004)]{ada04} Adamson, A., et al. 2004, in
Galactic Center Workshop 2002, The Central 300 parsecs of the Milky
Way, ed. A. Cotera, S. Markoff, T. R. Geballe, \& H. Falcke
(Weinheim: Wiley-VCH Verlag GmbH \& Co.), 211

\bibitem[Anderson et al.(1980)]{and80} Anderson, T. G., Gudeman, C.
S., Dixon, T. A., \& Woods, R. C. 1980, J. Chem. Phys., 72, 1332

\bibitem[Arimoto et al.(1996)]{ari96} Arimoto, N., Sofue, Y., \&
Tsujimoto, T. 1996, \pasj, 48, 275

\bibitem[Bally et al.(1987)]{bal87} Bally, J., Stark, A. A.,
Wilson, R. W., \& Henkel, C.  1987, \apjs, 65, 13

\bibitem[Bally et al.(1988)]{bal88} Bally, J., Stark, A. A.,
Wilson, R. W., \& Henkel, C. 1988, \apj, 324, 223

\bibitem[Becklin et al.(1978)]{bec78} Becklin, E. E., Matthews, K., Neugebauer,
G., \& Willner, S. P.  1978, \apj, 219, 121

\bibitem[Binney \& Tremaine(1987)]{bin87} Binney, J., \& Tremaine S.
1987, Galactic Dynamics (Princeton, Princeton University Press)

\bibitem[Binney et al.(1991)]{bin91} Binney, J., Gerhard, O. E., Stark, A. A., Bally, J.,
 \& Uchida, K. I. 1991, \mnras, 252, 210

\bibitem[Binney(1994)]{bin94} Binney, J., 1994, in NATO ASI Series, The Nuclei of Normal
Galaxies -- Lessons from the Galactic Center, ed. R. Genzel, \& A.
I. Harris (Dordrecht: Kluwer Academic Publishers), 75

\bibitem[Blitz et al.(1993)]{bli93} Blitz, L., Binney, J., Lo, K. Y., Bally, J.,
 \& Ho, P. T. P. 1993, Nature, 361, 417

\bibitem[Brittain et al.(2004)]{bri04} Brittain, S. D., Simon, T., Kulesa, C., \&
Rettig, T. W. 2004, \apj, 606, 911

\bibitem[Brown \& Liszt(1984)]{bro84} Brown, R. L., \& Liszt, H. S. 1984, \araa, 22,
223

\bibitem[Butchart et al.(1986)]{but86} Butchart, I., McFadzean, A. D., Whittet, D. C. B.,
Geballe, T. R., \& Greenberg, J. M. 1986, \aap, 154, L5

\bibitem[Cordonnier et al.(2000)]{cor00} Cordonnier, M., Uy, D., Dickson, R.
M., Kerr, K. E., Zhang, Y., \& Oka, T. 2000, J. Chem. Phys., 113,
3181

\bibitem[Cotera et al.(2000)]{cot00} Cotera, A. S., Simpson, J. P., Erickson, E.
F., Colgan, S. W. J., Burton, M. G., \& Allen, D. A. 2000, \apjs,
129, 123

\bibitem[Dahmen et al.(1998)]{dah98} Dahmen, G., H\"{u}ttemeister, S., Wilson, T.
L., \& Mauersberger, R. 1998, \aap, 331, 959

\bibitem[Figer et al.(1999)]{fig99} Figer, D. F., McLean, I. S., \&
Morris, M. 1999, \apj, 514, 202

\bibitem[Geballe et al.(1989)]{geb89} Geballe, T. R., Baas, F., \& Wade, R.
1989, \aap, 208, 255

\bibitem[Geballe \& Oka(1996)]{geb96} Geballe, T. R., \& Oka, T. 1996,
Nature, 384, 334

\bibitem[Geballe et al.(1999)]{geb99} Geballe, T. R., McCall, B. J., Hinkle, K. H., \& Oka, T. 1999,
\apj, 510, 251

\bibitem[Genzel et al.(1994)]{gen94} Genzel, R., Hollenbach, D., \& Townes, C. H. 1994,
Rep. Prog. Phys., 57, 417

\bibitem[Goto et al.(2002)]{got02} Goto, M., McCall, B. J., Geballe, T. R., Usuda, T., Kobayashi,
N., Terada, H., \& Oka, T. 2002, \pasj, 54, 951

\bibitem[Greaves(1995)]{gre95} Greaves, J. S. 1995, \mnras, 273, 918

\bibitem[Greaves et al.(1992)]{gre92} Greaves, J. S., White, G. J., Ohishi, M., Hasegawa, T.,
 \& Sunada, K. 1992, \aap, 260, 381

\bibitem[Green \& Chapman(1978)]{gre78} Green, S., \& Chapman, S. 1978, \apjs, 37,
169

\bibitem[G\"{u}sten \& Downes(1981)]{gus81a} G\"{u}sten, R., \& Downes,
D. 1981, \aap, 99, 27

\bibitem[G\"{u}sten et al.(1981)]{gus81} G\"{u}sten, R., Walmsley, C. M., \& Pauls, T.  1981,
 \aap, 103, 197

\bibitem[G\"{u}sten \& Henkel(1983)]{gus83} G\"{u}sten, R., \& Henkel, C.  1983, \aap, 125,
136

\bibitem[Harris et al.(1985)]{har85} Harris, A. I., Jaffe, D. T., Silber, M., \& Genzel, R.
1985, \apj, 294, L93

\bibitem[Hartquist et al.(1978a)]{har78a} Hartquist, T. W., Doyle, H. T., \& Dalgarno,
A. 1978, \aap, 68, 65

\bibitem[Hartquist et al.(1978b)]{har78b} Hartquist, T. W., Black, J. H., \& Dalgarno
A. 1978, \mnras, 185, 643


\bibitem[Herrnstein \& Ho(2002)]{her02} Herrnstein, R. M., \& Ho, P. T. P. 2002, \apj, 579, L 83

\bibitem[H\"{u}ttemeister et al.(1993)]{hut93} H\"{u}ttemeister, S., Wilson, T. L., Bania, T.
M., \& Mart\'{\i}n-Pintado, J. 1993, \aap, 280, 255

\bibitem[Kaifu et al.(1972)]{kai72} Kaifu, N., Kato, T., \& Iguchi, T.
1972, Nature, 238, 105

\bibitem[Kaifu et al.(1974)]{kai74} Kaifu, N., Iguchi, T., \& Kato, T.
1974, PASJ, 26, 117

\bibitem[Kim et al.(2002)]{kim02} Kim, S., Martin, C. L., Stark, A. A., \& Lane, A. P. 2002,
\apj, 580, 896

\bibitem[Kokoouline \& Greene(2003a)]{kok03a} Kokoouline, V., \& Greene, C. H. 2003a,
Phys. Rev. Lett., 90, 133201-1

\bibitem[Kokoouline \& Greene(2003b)]{kok03b} Kokoouline, V., \& Greene, C. H. 2003b,
Phys. Rev., A68, 012703

\bibitem[Koyama et al.(1989)]{koy89} Koyama, K., et al. 1989, Nature, 339, 603

\bibitem[Koyama et al.(1996)]{koy96} Koyama, K., Maeda, Y., Sonobe, T., Takeshima, T.,
Tanaka, Y., \& Yamauchi, S. 1996, \pasj, 48, 249

\bibitem[Kreckel et al.(2004)]{kre04} Kreckel, H., Tennyson, J.,
Schwalm, D., Zajfman, D., \& Wolf, A. 2004, New J. Phys., 6, 151

\bibitem[Kreckel et al.(2002)]{kre02} Kreckel, H., et al. 2002, Phys. Rev., A66, 052509

\bibitem[Lacy et al.(1994)]{lac94} Lacy, J. H., Knacke, R., Geballe, T. R., \&
Tokunaga, A. T. 1994, \apj, 428, L69

\bibitem[Lang et al.(2002)]{lan02} Lang, C. C., Goss, W. M., \& Morris, M. 2002,
\aj, 124, 2677

\bibitem[Lang et al.(2004)]{lan04} Lang, C. C., Cyganowski, C., Goss, W. M., \& Zhao, J.
-H. 2004, in Galactic Center Workshop 2002, The Central 300 parsecs
of the Milky Way, ed. A. Cotera, S. Markoff, T. R. Geballe, \& H.
Falcke (Weinheim: Wiley-VCH Verlag GmbH \& Co.), 1

\bibitem[Lee et al.(1996)]{lee96} Lee, H. -H., Bettens, R. P. A., \& Herbst, E. 1996,
\aaps, 119, 111

\bibitem[Le Petit et al.(2004)]{lep04} Le Petit, F., Roueff, E., \& Herbst, E. 2004,
\aaps, 417, 993

\bibitem[Liao \& Herbst(1996)]{lia96} Liao, Q., \& Herbst, E. 1996,
J. Chem. Phys., 104, 3956

\bibitem[Lindsay \& McCall(2001)]{lin01} Lindsay, C. M., \& McCall, B. J. 2001,
J. Mol. Spectrosc., 210, 60

\bibitem[Linke et al.(1981)]{lin81} Linke, R. A., Stark, A. A., \& Frerking, M. A.
1981, \apj, 243, 147

\bibitem[Liszt(2003)]{lis03} Liszt, H. S. 2003,
\aap, 398, 621

\bibitem[Liszt et al.(1985)]{lis85} Liszt, H. S., Burton, W. B., \& van der Hulst, J. M. 1985
\aap, 142, 237

\bibitem[Mauersberger et al.(1986)]{mau86} Mauersberger, R., Henkel, C., Wilson, T.
L., \& Walmsley, C. M. 1986, \aap, 162, 199

\bibitem[McCall et al.(1998a)]{mcc98a} McCall, B. J., Geballe, T. R., Hinkle, K. H., \& Oka, T.
1998a, Science, 279, 1910

\bibitem[McCall et al.(1999)]{mcc99} McCall, B. J., Geballe, T. R., Hinkle, K. H., \& Oka, T. 1999, \apj,
522, 338

\bibitem[McCall et al.(1998b)]{mcc98b} McCall, B. J., Hinkle, K. H., Geballe, T. R., \& Oka, T.
1998b, Faraday Discuss., 109, 267

\bibitem[McCall et al.(2002)]{mcc02} McCall, B. J., et al. 2002, \apj, 567, 391

\bibitem[McCall et al.(2003)]{mcc03} McCall, B. J., et al. 2003, Nature, 422, 500

\bibitem[McCall et al.(2004)]{mcc04} McCall, B. J., et al. 2004, Phys. Rev.,
A70, 052716

\bibitem[Menon \& Ciotti(1970)]{men70} Menon, T, K., \& Ciotti, J. E.
1970, Nature, 227, 579

\bibitem[Mezger et al.(1996)]{mez96} Mezger, P, G., Duschl, W. J., \& Zylka, R.
1996, Astron. Astrophys. Rev., 7, 289

\bibitem[Morris \& Serabyn(1996)]{mor96} Morris, M., \& Serabyn, E.
1996, \araa, 34, 645

\bibitem[Nagata et al.(1993)]{nag93} Nagata, T., Hyland, A. R., Straw, S. M., Sato, S., \&
Kawara, K.  1993, \apj, 406, 501

\bibitem[Nagata et al.(1990)]{nag90} Nagata, T., Woodward, C. E.,  Shure, M., Pipher, J. L., \& Okuda, H.
1990, \apj, 351, 83

\bibitem[Neale et al.(1996)]{nea96} Neale, L., Miller, S., \& Tennyson, J.  1996, \apj, 464, 516

\bibitem[Odenwald \& Fazio(1984)]{ode84} Odenwald, S. F., \& Fazio, G. G. 1984,
\apj, 283, 601

\bibitem[Oesterling et al.(2001)]{oes01} Oesterling, L. C., De Lucia, F. C. \& Herbst, E. 2001,
Spectrochim. Acta A, 57, 705


\bibitem[Oka(1980)]{oka80} Oka, T. 1980, Phys. Rev. Lett. 45, 531

\bibitem[Oka(2004a)]{Oka04a} Oka, T. 2004a, J. Mol. Spectrosc. 228,
635

\bibitem[Oka(2004b)]{oka04b} Oka, T. 2004b, in the 4th Cologne-Bonn-Zermatt-Symposium,
The Dense Interstellar Medium in Galaxies, ed. S. Pfalzner, C.
Kramer, C. Staubmeier, \& A. Heithausen (Heidelberg:
Springer-Verlag), 37

\bibitem[Oka \& Epp(2004)]{oke04} Oka, T., \& Epp, E.  2004, \apj, 613, 349,
Paper I

\bibitem[Oka et al.(1998b)]{oka98b} Oka, T., Hasegawa, T., Hayashi, M., Handa, T., \&
Sakamoto, S. 1998b, \apj, 493, 730

\bibitem[Oka et al.(1998a)]{oka98a} Oka, T., Hasegawa, T., Sato, F., Tsuboi, M., \&
Miyazaki, A. 1998a, \apjs, 118, 455

\bibitem[Oka et al.(2001)]{oka01} Oka, T., Hasegawa, T., Sato, F., Tsuboi, M., \&
Miyazaki, A. 2001, \pasj, 53, 779

\bibitem[Oka et al.(1971)]{oka71} Oka, T., Shimizu, F. O., Shimizu, T., \&
Watson, J. K. G. 1971, \apj, 165, L15

\bibitem[Okuda et al.(1990)]{oku90} Okuda, H., et al. 1990, \apj, 351, 89

\bibitem[Oort(1977)]{oor77} Oort, J. H. 1977, \araa, 15, 295

\bibitem[Pan \& Oka(1986)]{pan86} Pan, F.-S., \& Oka, T.  1986, \apj, 305, 518

\bibitem[Pearson et al.(1995)]{pea95} Pearson, J. C., Oesterling, L. C., Herbst, E., \& De Lucia, F. C.
1995, Phys. Rev. Lett. 75, 2940

\bibitem[Pendleton et al.(1994)]{pen94} Pendleton, Y. J., Sanford S. A., Allamandola, L.
J., Tielens, A. G. G. M., \& Sellgren, K. 1994, \apj, 437, 683

\bibitem[Pierce-Price et al.(2000)]{pie00} Pierce-Price, D., et al. 2000,
\apj, 545, L121

\bibitem[Ridge (2004)]{rid04} Ridge, D. P. 2004, in
The Encyclopedia of Mass Spectrometry, Vol. 1, Theory and Ion
Chemistry, ed. P. B. Armentrout (Amsterdam: Elsevier) 1

\bibitem[Rodr\'{\i}guez-Fern\'{a}ndez et al.(2001)]{rod01}
Rodr\'{\i}guez-Fern\'{a}ndez, N. J., Mart\'{\i}n-Pintado, J.,
Fuente, A., de Vincente, P., Wilson, T. L., \& H\"{u}ttemeister,
S. 2001, \aap, 365, 174

\bibitem[Rodr\'{\i}guez-Fern\'{a}ndez et al.(2004)]{rod04}
Rodr\'{\i}guez-Fern\'{a}ndez, N. J., Mart\'{\i}n-Pintado, J.,
Fuente, A., \& Wilson, T. L. 2004, \aap, 427, 217

\bibitem[Rougoor \& Oort(1960)]{rou60} Rougoor, G. W., \& Oort, J. H. 1960, Proc. Natl. Acad. Sci.
USA, 46, 1

\bibitem[Sawada et al.(2001)]{saw01} Sawada, T., et al. 2001, \apjs, 136, 189

\bibitem[Scoville(1972)]{sco72} Scoville, N. Z. 1972, \apj,
175, L127

\bibitem[Sofue(1995)]{sof95} Sofue, Y. 1995, \pasj, 47, 551

\bibitem[Sodroski et al.(1995)]{sod95} Sodroski, T. J., et al.
1995, \apj, 452, 262

\bibitem[Sofia et al.(2004)]{sof04} Sofia, U. J., Lauroesch, J. T., Meyer, D. M., \&
Cartledge, S. I. B. 2004, \apj, 605, 272

\bibitem[Tollestrup et al.(1989)]{tol89} Tollestrup, E. V., Capps, R. W., \&
Becklin, E. E. 1989, \aj, 98, 204

\bibitem[Tsuboi et al.(1997)]{tsu97} Tsuboi, M., Ukita, N., \& Handa, T.
 1997, \apj, 481, 263

\bibitem[Tsuboi et al.(1999)]{tsu99} Tsuboi, M., Handa, T., \&
Ukita, N. 1999, \apjs, 120, 1

\bibitem[van Dishoeck \& Black(1986)]{van86} van Dishoeck, E. F., \&
Black, J. H. 1986, \apjs, 62, 109

\bibitem[van Woerden et al.(1957)]{van57} van Woerden, H., Rougoor, G. W., \&
Oort, J. H. 1957, C. R. Acad. Sciences, Paris, 244, 1691

\bibitem[Wada et al.(1994)]{wad94} Wada, K., Taniguchi, Y., Habe, A., \&
Hasegawa, T. 1994, \apj, 437, L 123

\bibitem[Welty \& Hobbs(2001)]{wel01} Welty, D. E., \& Hobbs, L. M. 2001, \apjs, 133,
345

\bibitem[Whiteoak \& Gardner(1979)]{whi79} Whiteoak, J. B., \&
Gardener, F. F. 1979, \mnras, 188, 445

\bibitem[Whittet et al.(1997)]{whi97} Whittet, D. C. B., et al. 1997, \apj, 490, 729

\bibitem[Willner et al.(1979)]{wil79} Willner, S. P., Russell, R. W., Puetter, R. C., Soifer, B. T.,
\& Harvey, P. M. 1979, \apj, 229, L65

\bibitem[Wilson et al.(1982)]{wil82} Wilson, T. L., Ruf, K., Walmsley, C. M., Martin, R. N.,
Pauls, T. A., \& Batrla, W. 1982, \aap, 115, 185

\bibitem[Wright et al.(2001)]{wri01} Wright, M. C. H., Coil, A. L., McGray, R. S., Ho, P. T. P.,
\& Harris, A. I. 2001, \apj, 551, 254

\bibitem[Yusef-Zadeh et al.(2002)]{yus02} Yusef-Zadeh, F., Law, C.,
\& Wardle, M. 2002, \apj, 568, L121

\bibitem[Yusef-Zadeh \& Morris(1987)]{yus87} Yusef-Zadeh, F., \& Morris, M. 1987,
\aj, 94, 1178

\bibitem[Yusef-Zadeh et al.(1984)]{yus84} Yusef-Zadeh, F., Morris, M.,
\& Chance, D. 1984, Nature, 310, 557

\end{thebibliography}
\end{document}